\documentclass[nofootinbib,aps,12pt]{revtex4-1}
\usepackage{graphicx}
\usepackage{hyperref}
\usepackage{cancel}
\usepackage{amssymb}
\usepackage{amsmath}
\usepackage{amsfonts}
\usepackage{bm}
\usepackage{times}
\usepackage{color}
\usepackage{mathptmx}
\usepackage[normalem]{ulem}
\usepackage{ mathrsfs }
\usepackage{textcomp}
\usepackage{slashed}
\usepackage{booktabs}
\usepackage{latexsym}
\usepackage{verbatim}
\usepackage{extarrows}
\usepackage{multirow}
\usepackage{xcolor}
\usepackage{epsfig}
\usepackage{epstopdf}
\usepackage{graphics}
\usepackage{subfigure,psfrag}
\hypersetup{
	colorlinks,
	linkcolor={black!50!black},
	citecolor={blue!50!black},
	urlcolor={blue!80!black},
	bookmarksopen=true
}
\makeatletter
\renewcommand*\env@matrix[1][\arraystretch]{%
	\edef\arraystretch{#1}%
	\hskip -\arraycolsep
	\let\@ifnextchar\new@ifnextchar
	\array{*\c@MaxMatrixCols c}}
\makeatother
\begin{document}

\title{Electroweak phase transition and Higgs phenomenology in the Georgi-Machacek model}

\author{Ruiyu Zhou $^{1}$}

\author{Wei Cheng $^{1}$}

\author{Xin Deng $^{1}$}

\author{Ligong Bian $^{1,2}$}
\email{lgbycl@cqu.edu.cn}

\author{Yongcheng Wu$^{3}$}
\email{ycwu@physics.carleton.ca}

\affiliation{
	$^1$~Department of Physics, Chongqing University, Chongqing 401331, China
	\\
    $^2$ Department of Physics, Chung-Ang University, Seoul 06974, Korea\\
    $^3$ Ottawa-Carleton Institute for Physics, Carleton University, 1125 Colonel By Drive, Ottawa,
Ontario K1S 5B6, Canada
}
\date{\today}

\begin{abstract}

In this work, we perform the electroweak phase transition study with the Georgi-Machacek model. We investigate both the one-step and two-step
strong first order electroweak phase transition (SFOEWPT). The SFOEWPT viable parameter spaces could be tested by the future 14 TeV LHC, HL-LHC, and ILC.  The LHC Higgs signal strength measurements severely bound the SFOEWPT valid parameter spaces, a tinny region of the mixing angle between the neutral fields of the isospin-doublet and isospin-triplet scalars around $\alpha\sim 0$ can allow the two-step SFOEWPT to occur. The triplet vacuum expectation value (VEV) is crucial for both SFOEWPT and related Higgs phenomenology. The two-step SFOEWPT can be distinguished from the one-step SFOEWPT through the triple Higgs coupling searches and the low mass doubly charged Higgs searches at colliders.

\end{abstract}

\graphicspath{{figure/}}

\maketitle
\baselineskip=16pt

\pagenumbering{arabic}

\vspace{1.0cm}
\tableofcontents

\newpage

\section{Introduction}
\label{sec:intro}
The baryon asymmetry of the Universe (BAU) as a fundamental problem of the nature has puzzled particle and cosmology physicists for several decades. To address the problem, three Sakharov conditions are necessary~\cite{Sakharov:1967dj}. Wherein, the CP violation and the departure from thermodynamic equilibrium all call for new physics beyond the SM. Among variant baryogenesis mechanisms, the electroweak baryogenesis mechanism (EWBG) inspires people interest recently due to the two key ingredients that are testable~\cite{Morrissey:2012db}: One crucial ingredient is the CP violation beyond the SM that bias the sphalerons to create baryon asymmetry, it could be probed indirectly with the high precision
electric dipole moment experiments (EDMs).
 Another crucial ingredient is the strongly first order electroweak phase transition (SFOEWPT) that prevents the sphaleron process washing out the baryon asymmetry, which naturally provide an explanation of symmetry breaking as the Universe temperature cools down and could be tested at colliders~\cite{Arkani-Hamed:2015vfh}.

 In this work, we will focus on the SFOEWPT study where the Higgs phenomenology are more involved.
The generated gravitational wave signals from which with a typical peak frequency O(10$^{-3}$ - 10$^{-1}$)Hz is detectable at
the projected space-based interferometers, such as:  eLISA~\cite{Caprini:2015zlo}, BBO, DECIGO (Ultimate-DECIGO)~\cite{Kudoh:2005as} and ALIA~\cite{Gong:2014mca}. After the gravitational waves signals from the merging black holes are detected by LIGO ~\cite{Abbott:2016blz}, the gravitational waves from the SFOEWPT has inspired extensive studies in various BSM models with the SM Higgs sectors being extended.
To achieve SFOEWPT, it is well known that the scalar sectors of the SM should be extended and can lead to testable signals at colliders~\cite{Arkani-Hamed:2015vfh}, including the $e^+e^-$ colliders (such as the projected CEPC~\cite{CEPCStudyGroup:2018ghi}, ILC~\cite{Baer:2013cma}, and FCC-ee~\cite{Gomez-Ceballos:2013zzn}) and high luminosity pp colliders (such as FCC-hh~\cite{FCC,FCChh,CERN} and SPPC~\cite{CEPCStudyGroup:2018ghi}).
 On the other hand, in the present post-Higgs era, the ongoing collider search of heavy and charged Higgs beyond the SM also assumes the extension of the SM scalar sectors.

Among various extensions of the SM, the Georgi-Machacek model introduces a real and a complex isospin-triplet scalar to extend the SM Higgs sector~\cite{Georgi:1985nv,Chanowitz:1985ug}, which introduce a new contribution of the electroweak symmetry breaking. The custodial symmetry there is preserved explicitly at tree level by assuming the same vacuum expectation value (VEV) to the neutral fields of the two isospin-triplet scalars~\cite{Hartling:2014zca}.
The interaction strength of the SM vector bosons and the doubly- and singly-charged Higgs bosons of the quintuple is controlled by the vacuum expectation value (VEV) of the triplets, which can be probed through the charged Higgs searches~\cite{Chiang:2012cn,Chiang:2013rua,Logan:2017jpr}.
A recent study on the heavy Higgs and charged Higgs in the ``H5plane" benchmark scenario for the GM model (developed by the LHC Higgs Cross Section Working Group~
\cite{deFlorian:2016spz} ) can be found in Ref.~\cite{Logan:2017jpr}.
The triplet scalars can introduce additional vacuum structures (local minimum), which might make the symmetry breaking history occur through a two-step pattern since the $SU(2)_L\times SU(2)_R\to SU(2)_V$ can occur before the electroweak symmetry breaking. Therefore, we perform the study of phase transition patterns to find the relation between the SFOEWPT conditions and the Higgs phenomenology. As for the relation between the SFOEWPT and the charged Higgs bosons phenomenology, we focus on the low mass benchmark not yet covered by the LHC~\cite{Logan:2018wtm}.
This constitutes our motivation to study electroweak phase transition within the GM model and makes our study significantly different from the previous one~\cite{Chiang:2014hia}.
In this work we focus on the SFOEWPT study utilizing the custodial vacuum alignment scenario of the GM model. The CP violation of the GM model requires custodial symmetry breaking, we left that to the future work.

This work is organized as follows: We first review the GM model under study in Sec.~\ref{sec:mod}. The electroweak phase transition methodology being adopted is explored in Sec.~\ref{sec:ewptm}. The interplay between the SFOEWPT condition and the related Higgs phenomenology are constructed in Sec.~\ref{sec:ptph}. The Sec.~\ref{sec:conc} is attributed to the conclusions.

\section{The Georgi-Machacek Model}
\label{sec:mod}

\subsection{The Model Setup}
In the Georgi-Machacek model, there are one isospin doublet scalar field $\phi = (\phi^+,\phi^0)^T$ with hypercharge $\rm Y=\frac{1}{2}$, one complex isospin triplet scalar field $\chi = (\chi^{++},\chi^+,\chi^0)^T$ with hypercharge $\rm Y=1$, and one real triplet $\xi = (\xi^+,\xi^0,-\xi^{+*})^T$ with hypercharge $\rm Y=0$. The custodial symmetry is introduced at tree level by imposing a global SU(2)$_L\times$SU(2)$_R$ symmetry upon the scalar potential. In order to make this symmetry explicit,
the doublet and the two triplets are written in the form of a bi-doublet and a bi-triplet under SU(2)$_L\times$SU(2)$_R$:
\begin{align}
\label{equ:scalar_components}
\Phi \equiv \left(\epsilon_2\phi^*, \phi \right) = \left(\begin{array}{cc}
\phi^{0*} & \phi^+ \\
-\phi^{+*} & \phi^0
\end{array}\right), \quad \Delta \equiv \left(\epsilon_3\chi^*,\xi,\chi\right) = \left(\begin{array}{ccc}
\chi^{0*} & \xi^+ & \chi^{++} \\
-\chi^{+*} & \xi^0 & \chi^+ \\
\chi^{++*} & -\xi^{+*} & \chi^0
\end{array}\right)\;,
\end{align}
with
\begin{align}
\epsilon_2 = \left(\begin{array}{cc}
0 & 1 \\
-1 & 0
\end{array}\right),\quad \epsilon_3 = \left(\begin{array}{ccc}
0 & 0 & 1 \\
0 & -1 & 0\\
1 & 0 & 0
\end{array}\right)\;,
\end{align}
where the phase convention for the scalar field components is:
 $\chi^{--}=\chi^{++*},\,\, \chi^{-}=\chi^{+*},\,\, \xi^-= \xi^{+*},\,\, \phi^-= \phi^{+*}~$. $\Phi$ and $\Delta$ are transformed under $SU(2)_L \times SU(2)_R$ as $\Phi \to U_{2,L} \Phi U_{2,R}^\dagger$ and $\Delta \to U_{3,L} \Delta U_{3,R}^\dagger$ with $U_{L,R}=exp(i\theta^a_{L,R} T^a)$ and $T^a$ being the $SU(2)$ generators.


The most general scalar potential $V(\Phi, \, \Delta)$ invariant under $SU(2)_L \times SU(2)_R \times U(1)_Y$ is given by
\begin{align}
V(\Phi, \Delta)=& \frac{1}{2} m_1^2 {\rm tr}[ \Phi^{\dagger} \Phi ] +
\frac{1}{2} m_2^2 {\rm tr}[ \Delta^{\dagger} \Delta ]
 +  \lambda_1 \left( {\rm tr}[ \Phi^{\dagger} \Phi ] \right)^2 \nonumber\\
& +  \lambda_2 \left( {\rm tr}[ \Delta^{\dagger} \Delta ] \right)^2
 +  \lambda_3 {\rm tr}\left[ \left( \Delta^{\dagger} \Delta \right)^2 \right]
+  \lambda_4 {\rm tr}[ \Phi^{\dagger} \Phi ] {\rm tr}[ \Delta^{\dagger} \Delta ] \nonumber\\
& +  \lambda_5 {\rm tr}\left[ \Phi^{\dagger} \frac{\sigma^a}{2} \Phi \frac{\sigma^b}{2} \right]
                  {\rm tr}[ \Delta^{\dagger} T^a \Delta T^b] \nonumber \\
 &+ \mu_1 {\rm tr}\left[ \Phi^{\dagger} \frac{\sigma^a}{2} \Phi \frac{\sigma^b}{2} \right]
                               (P^{\dagger} \Delta P)_{ab}
 + \mu_2 {\rm tr}[ \Delta^{\dagger} T^a \Delta T^b]
                               (P^{\dagger} \Delta P)_{ab} ~,
\label{potential}
\end{align}
where summations over $a,b = 1,2,3$ are understood, $\sigma$'s and $T$'s are the $2\times2$ (Pauli matrices) and $3\times3$ matrix representations of the $SU(2)$ generators, respectively
\begin{align}
T_1 = \frac{1}{\sqrt{2}} \left( \begin{array}{ccc}
0 & 1 & 0 \\
1 & 0 & 1 \\
0 & 1 & 0
\end{array}
\right),
T_2 &= \frac{1}{\sqrt{2}} \left( \begin{array}{ccc}
0 & -i & 0 \\
i & 0 & -i \\
0 & i & 0
\end{array}
\right),
T_3 = \left( \begin{array}{ccc}
1 & 0 & 0 \\
0 & 0 & 0 \\
0 & 0 & -1
\end{array}
\right),
\end{align}
The P matrix, which is the similarity transformation relating the generators in the triplet and the adjoint representations, is given by
\begin{align}
P &= \frac{1}{\sqrt{2}} \left( \begin{array}{ccc}
-1 & i & 0 \\
0 & 0 & \sqrt{2} \\
1 & i & 0
\end{array}
\right)\;.
\end{align}
The neutral components in Eq.(\ref{equ:scalar_components}) can be parameterized into real and imaginary parts according to
\begin{eqnarray*}
\rm \phi^0 = \frac{\nu_{\phi} + h_\phi + i a_\phi}{\sqrt{2}},
\rm \qquad
\rm \chi^0 = \frac{\nu_{\chi} + h_\chi + i a_\chi}{\sqrt{2}},
\rm \qquad
\rm \xi^0 = \nu_{\xi} + h_\xi,
\end{eqnarray*}
where $\nu_\phi$, $\nu_\chi$ and $\nu_\xi$ are the VEVs of $\phi^0$, $\chi^0$ and $\xi^0$, respectively.
With only neutral components of this model, the potential reads:
\begin{eqnarray}
V_{0}&=& \frac{1}{4}(4 h_{\phi}^{4} \lambda_1 + 2 (h_{\xi}^2+h_{\chi}^2)(m_2^2+2(h_{\xi}^2+h_{\chi}^2)\lambda_2)+2\lambda_3(2 h_{\xi}^4 + h_{\chi}^4)\nonumber\\
&+& h_{\phi}^2 (2 m_1^2 + 4 h_{\xi}^2 \lambda_4 + h_{\xi}(2\sqrt{2} h_{\xi} \lambda_5 + \mu_1) + h_{\xi}(4 h_{\xi} \lambda_4 + h_{\xi} \lambda_5 + \sqrt{2} \mu_1)) + 12 h_{\xi} h_{\chi}^2 \mu_2 ).\label{eq:Vtree}
\end{eqnarray}
We can derived the EWSB vacuum from this conditions:
\begin{eqnarray}
\frac{\partial V_{0}}{\partial h_{\phi}} \ = \ \frac{\partial V_{0}}{\partial h_{\chi}}
\ = \ \frac{\partial V_{0}}{\partial h_{\xi}} \ = \ 0 ~,
\label{vacuum}
\end{eqnarray}
where the fields other than
$\phi^0$, $\chi^0$ and $\xi^0$
take zero VEV's. In this paper, the solution satisfying the relation $v_{\chi}=\sqrt{2}v_{\xi}$ is selected, by which the EWSB vacuum maintains the diagonal $SU(2)_V$ symmetry. Thus the parameter $\rho = m_W^2/(m_Z^2 \cos\theta_w^2) = 1$ is established at the tree level.
The W and Z boson masses from the EWSB give the constraint,
\begin{equation}
\nu_\phi^2 + 8\nu_\xi^2 \equiv \nu^2 = \frac{1}{\sqrt{2}G_F}\approx (246\text{ GeV})^2\;.
\end{equation}

When $\nu_\phi$ $\nu_\xi \neq 0$, with the help of Eq.({\ref{vacuum}}) (under the relation $\nu_\chi = \sqrt{2}\nu_\xi$), we could rewrite $m_1^2$, $m_2^2$ in terms of $\nu_\phi$,$\nu_\xi$ and other parameters in the Higgs potential as:
\begin{align}
m_1^2 \ &= \ -4\lambda_1 \nu_{\phi}^2 - 6\lambda_4 \nu_\xi^2 - 3\lambda_5 \nu_\xi^2
- \frac{3}{2} \mu_1 \nu_\xi \;,\\
m_2^2 \ &= \ -12\lambda_2 \nu_\xi^2 - 4\lambda_3 \nu_\xi^2 - 2\lambda_4 \nu_{\phi}^2
- \lambda_5 \nu_{\phi}^2 - \mu_1 \frac{\nu_{\phi}^2}{4\nu_\xi} - 6\mu_2 \nu_\xi\;.
\label{m1m2}
\end{align}
There are 13 scalar fields in this model.
After diagonalizing the mass matrices, the fields can be rewritten as the physical scalars (quintuple, triplet and singlet respectively)
\begin{align}
& H_5^{++} = \chi^{++} ~,\quad H_5^+ = \frac{1}{\sqrt{2}} \bigg( \chi^+ - \xi^+ \bigg) ~,\quad  H_5^0 = \sqrt{\frac{1}{3}} h_{\chi} - \sqrt{\frac{2}{3}} h_{\xi} ~,\\
&H_3^+ = -\cos \theta_H \, \phi^+ + \sin \theta_H \, \frac{1}{\sqrt{2}} \bigg( \chi^+ + \xi^+ \bigg) ~,
\quad  H_3^0 = -\cos \theta_H \, a_{\phi} + \sin \theta_H \, a_{\chi} ~, \\
&h = \cos \alpha \, h_{\phi} - \frac{\sin \alpha}{\sqrt3} \, \bigg( \sqrt{2} h_{\chi} + h_{\xi} \bigg) ~,\quad H_1 = \sin \alpha \, h_{\phi} + \frac{\cos \alpha}{\sqrt3} \, \bigg( \sqrt{2} h_{\chi} + h_{\xi} \bigg) ~,
\end{align}
and the goldstone bosons
\begin{align}
& G^+ = \sin\theta_H \phi^+ + \cos\theta_H \frac{1}{\sqrt2}(\chi^+ + \xi^+)~, ~~G^0 = \sin\theta_H a_\phi + \cos\theta_H a_\xi\;,
\end{align}
where $\sin\theta_H = \frac{2\sqrt{2}\nu_\xi}{\nu}$ and $\cos\theta_H = \frac{\nu_\phi}{\nu}$, and $\alpha$ is the mixing angle between two singlets which is determined by the mass matrix of these scalars as will be shown below.

The 3 goldstone bosons eventually become the longitudinal components of the W and Z bosons, while, the remaining 10 physical fields can be organized into a quintuple $H_5$ $=$ $(H^{++}_5$, $H^+_5$, $H^0_5$, $H^-_5$, $H^{--}_5)^T$, a triplet $H_3 = (H^+_3, H^0_3, H^-_3)^T$ and two singlets $h$ and $H_1$, where the former ($h$) is used to denote the SM-like Higgs boson. The triplet scalar is CP-odd, while others are CP-even.
The masses of different multiplets can be written as
\begin{align}
m^2_{H_5}=&m^2_{H_5^{\pm\pm}}=m^2_{H_5^{\pm}}=m^2_{H_5^0} = (8 \lambda_3 \nu^2_{\xi}- \frac{3}{2}\lambda_5 \nu^2_{\phi})  - \frac{\mu_1 \nu^2_{\phi}}{4 \nu_\xi} - 12 \mu_2\nu_\xi \;,\\
m^2_{H_3}=&m^2_{H_3^\pm}=m^2_{H_3^0} = - (\frac{\lambda_5}{2} + \frac{\mu_1}{4\nu_\xi})\nu^2\;.
\end{align}
The singlets masses of $m_{h,H_1}$ are the eigenvalues of mass matrix written in terms of gauge eigenstates:
\begin{align}
M^2 = \left(\begin{array}{cc}
M^2_{11} & M^2_{12} \\
M^2_{12} & M^2_{22}
\end{array}\right)
\end{align}
with
\begin{align}
M^2_{11} &= 8 \cos^2\theta_H \lambda_1 \nu^2,\\
M^2_{22} &= \sin^2\theta_H (3\lambda_2+\lambda_3)\nu^2 + \cos^2\theta_H M^2_1-\frac{1}{2}M_2^2, \\
M^2_{12} &= \sqrt{\frac{3}{2}}\sin\theta_H \cos\theta_H[(2\lambda_4+\lambda_5)\nu^2-M_1^2],
\end{align}
where $M_1^2 = -\frac{\nu}{\sqrt{2}\sin\theta_H}\mu_1$ and $M_2^2=-3\sqrt{2}\sin\theta_H \nu \mu_2$. The mixing angle $\alpha$ is determined by
\begin{equation}
\tan2\alpha = \frac{2M^2_{12}}{M^2_{11}-M^2_{22}}\;.
\end{equation}
The five dimensionless couplings in the potential, $\lambda_{1,2,3,4,5}$, can be substituted by five physical parameters $m_{H_1},~m_{H_3},~m_{H_5},~\alpha,~\theta_H$.
\begin{align}
\lambda_1 =&\frac{1}{8v^2\cos^2\theta_H}(m_h^2\cos^2\alpha+m_{H_1}^2\sin^2\alpha),\notag\\
\lambda_2 =&\frac{1}{6v^2\sin^2\theta_H}
[2m_{H_1}^2\cos^2\alpha+2m_h^2\sin^2\alpha+3M_2^2-2m_{H_5}^2+6\cos^2 \theta_H(m_{H_3}^2-M_1^2)],\notag\\
\lambda_3 =& \frac{1}{v^2\sin^2\theta_H}\left[\cos^2\theta_H(2M_1^2-3m_{H_3}^2)+m_{H_5}^2-M_2^2\right],\notag\\
\lambda_4=& \frac{1}{6v^2\sin\theta_H \cos\theta_H}\left[\frac{\sqrt{6}}{2}\sin2\alpha(m_h^2-m_{H_1}^2)+3\sin\theta_H \cos\theta_H(2m_{H_3}^2-M_1^2)\right],\notag\\
\lambda_5=& \frac{2}{v^2}(M_1^2-m_{H_3}^2). \label{lambdas}
\end{align}

\subsection{The Theoretical Constraints}
Three theoretical constraints are taken into account to constrain the dimensionless quartic couplings of the scalar potential at tree level: the unitarity of the perturbation theory, the stability of the electroweak vacuum and avoiding custodial symmetry-breaking vacuum. All these constraints have been investigated in detail in Ref.~\cite{Hartling:2014zca}, and will be automatically imposed on our parameter scan using GMCalc~\cite{Hartling:2014xma}.
\subsubsection{Tree-level unitarity}
The bound from perturbative unitarity is obtained by requiring that the zeroth partial
wave amplitude, $a_0$, for elastic $ 2 \to 2 $ scalar boson scatterings does not become too large to violate S-matrix unitarity. That means that the amplitude $a_0$ satisfy $|a_0| \leq 1$ or $|Re a_0|\leq 1/2$.
Then, the perturbative unitarity bound reads:
\begin{align}
\begin{split}
&
\left\vert \, 6 \lambda_1 + 7 \lambda_3 + 11\lambda_2\, \,\right\vert  + \sqrt{(6\lambda_1-7\lambda_3-11\lambda_2)^2+36\lambda_4^2} < 4\pi \;,
\\
&
\left\vert \, 2 \lambda_1 -  \lambda_3 + 2\lambda_2\, \,\right\vert  + \sqrt{(2\lambda_1+\lambda_3-2\lambda_2)^2+\lambda_5^2}  < 4\pi \;,
\\
&
\left\vert \,  \lambda_4 + \lambda_5 \, \right\vert < 2\pi \;,
~~~ \left\vert \, 2 \lambda_3 + \lambda_2  \, \right\vert < \pi ~\;,
\\
&\left\vert \,  2\lambda_2 + \lambda_3 \, \right\vert < 2\pi \;,
~~~ \left\vert \, 4 \lambda_4 + \lambda_5  \, \right\vert < 8\pi \;,~~~ \left\vert \, 2 \lambda_4 - \lambda_5  \, \right\vert < 4\pi \;.
\end{split}
\label{eq:unitarity}
\end{align}


\subsubsection{Vacuum stability constraints}
Following the approach of Ref.~\cite{Arhrib:2011uy}, we can parameterize the potential using the following definitions:
\begin{align}
	&r \equiv  \sqrt{\text{Tr}(\Phi^\dagger \Phi) + \text{Tr}(X^\dagger X)} ,\nonumber \\
    &r^2 \cos^2 \gamma  \equiv  \text{Tr}(\Phi^\dagger \Phi), \quad r^2 \sin^2 \gamma \equiv   \text{Tr}(X^\dagger X) ,\nonumber \\
    &\zeta \equiv   \frac{ \textrm{Tr}(X^\dagger X X^\dagger X)}{[{\rm Tr}(X^\dagger X)]^2},  \nonumber \\
	&\omega \equiv   \frac{\text{Tr}( \Phi^\dagger \tau^a \Phi \tau^b) \text{Tr}( X^\dagger t^a X t^b)}{\textrm{Tr}(\Phi^\dagger \Phi)\textrm{Tr}(X^\dagger X)}, \nonumber \\
 &\sigma \equiv \frac{{\rm Tr}(\Phi^{\dagger} \tau^a \Phi \tau^b) (U X U^{\dagger})_{ab}}
	{{\rm Tr}(\Phi^{\dagger} \Phi) [ {\rm Tr}(X^{\dagger} X)]^{1/2}}, \nonumber \\
	&\rho \equiv \frac{{\rm Tr}(X^{\dagger} t^a X t^b) (U X U^{\dagger})_{ab}}
	{[{\rm Tr}(X^{\dagger} X)]^{3/2}}.
	\label{eq:bfbdefs}
\end{align}
The quartic terms in the potential are given in this parametrization by,
\begin{align}
V=\frac{r^4}{(1+\tan^2 \gamma)^2}
\left[ \lambda_1 + (\lambda_4 + \omega \lambda_5) \tan^2 \gamma +
(\zeta \lambda_3 + \lambda_2 ) \tan^4 \gamma \right]
\end{align}
The vacuum stability requires the scalar potential to be bounded from below and leads to following constraints on the quartic couplings Ref.~\cite{Chiang:2018xpl}.
\begin{align}
\begin{split}
&\lambda_1>0~,\\
&\lambda_2>
\left\{
\begin{array}{l}
-\frac{1}{3}\lambda_3 \ \text{ for }\lambda_3\geq0~,
\\
-\lambda_3  \ \ \ \text{ for }\lambda_3<0~,
\end{array}
\right.\\
&\lambda_4>
\left\{
\begin{array}{l}
-\frac{1}{2}\lambda_5-2\sqrt{\lambda_1(\frac{1}{3}\lambda_3+\lambda_2)} \qquad\qquad
\text{ for }\lambda_5\leq0\text{ and }\lambda_3\geq0~,
\\
-\omega_+(\zeta)\lambda_5-2\sqrt{\lambda_1(\zeta\lambda_3+\lambda_2)}  \qquad
\text{ for }\lambda_5\leq0\text{ and }\lambda_3<0~,\\
-\omega_-(\zeta)\lambda_5-2\sqrt{\lambda_1(\zeta\lambda_3+\lambda_2)}  \qquad
\text{ for }\lambda_5>0~,
\end{array}
\right.
\end{split}
\label{eq:stability}
\end{align}
where
\begin{align}
&\omega_\pm(\zeta)=\frac{1}{6}(1-B)\pm\frac{\sqrt2}{3}\left[\left(1-B\right)\left(\frac{1}{2}+B\right)\right]^{1/2}~,\\
& B(\zeta) \equiv \sqrt{\frac{3}{2}\left[\zeta-\frac{1}{3}\right]}\in[0,1],~~~\zeta\in[\frac{1}{3},1]\;.
\end{align}

\subsubsection{Absence of deeper custodial symmetry-breaking minima}
With the notation in Eq.(\ref{eq:bfbdefs}), the full scalar potential can be written as:
\begin{eqnarray}
	V &=& \frac{r^2}{(1 + \tan^2 \gamma)} \frac{1}{2} \left[ m_1^2 + m_2^2 \tan^2 \gamma \right]
	\nonumber \\
	&& + \frac{r^4}{(1+\tan^2 \gamma)^2}
	\left[ \lambda_1 + (\lambda_4 + \omega \lambda_5) \tan^2 \gamma +
	(\zeta \lambda_3 + \lambda_2 ) \tan^4 \gamma \right]
	\nonumber \\
	&& + \frac{r^3}{(1 + \tan^2\gamma)^{3/2}}  \tan\gamma \left[ \sigma \mu_1+
	 \rho \mu_2 \tan^2 \gamma \right],\nonumber\\
	\label{eq:Valtmin}
\end{eqnarray}
For a check that the scalar potential possesses no custodial symmetry-breaking minima that are deeper than the desired custodial symmetry-preserving minimum, we could parameterize these values as below:
\begin{align}
\zeta &= \frac{1}{2}\sin^4\theta+\cos^4 \theta, ~~~\omega = \frac{1}{4}\sin^2\theta+\frac{1}{\sqrt{2}}\sin\theta \cos\theta, \nonumber\\
\sigma&= \frac{1}{2\sqrt{2}} \sin\theta + \frac{1}{4} \cos\theta, ~~~\rho = 3\sin^2\theta \cos \theta.
\end{align}
Our desired electroweak-breaking and custodial $SU(2)$-preserving vacuum corresponds to $\theta = \cos^{-1}(1/\sqrt{3})$, thus $\zeta = 1/3$ and $\omega_+ = 1/2$. The vacuum $\theta = \pi+a$ is also acceptable; it corresponds to a negative $\nu_\chi$. Other values of $\theta$ correspond to vacua that spontaneously break custodial $SU(2)$.

\section{Electroweak Phase transition methodology}
\label{sec:ewptm}

With the temperature cooling down, the Universe can evolve from the symmetric phase to the symmetry broken phase. The critical behavior can be studied with the finite temperature effective potential with particle physics models, through which one can obtain the critical classics field value and temperature being $v_C$ and $T_C$.
 Roughly speaking, SFOEWPT can be obtained when $v_C/T_C > 1$, with which, one may have the
electroweak sphaleron process quenched inside the bubble after bubble nucleation, and therefore
prevent the washout of the baryon asymmetry generated within the EWBG mechanism~\cite{Morrissey:2012db}.
We first analyse the vacuum structure of the potential at zero temperature, then demonstrate the phase transition computation method adopted in this work.

\subsection{Vacuum structure analysis}

\begin{figure*}[!htp]
\begin{center}
\includegraphics[width=0.4\textwidth]{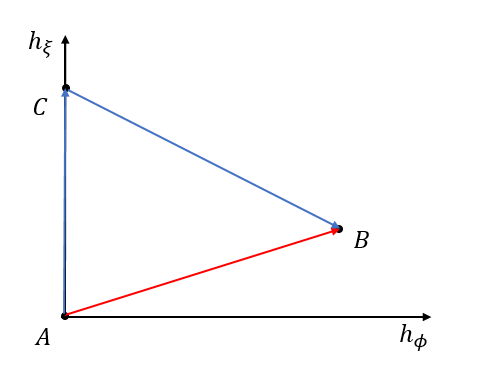}
\end{center}
\caption{One-step (red) and  two-step phase (blue) transition process. A is the original point, which is $SU(2)_L \times SU(2)_R$ preserving vacuum. B is the EW vacuum and C is the $SU(2)_V$ preserving vacuum.}\label{fig:PTpattern}
\end{figure*}

Assume $h_\chi = \sqrt{2} h_\xi$, the vacuum structure of the zero temperature potential could be obtained through the following minimization conditions
\begin{align}
\frac{\partial V_{0}}{\partial h_{\phi}} \ = \ \frac{\partial V_{0}}{\partial h_{\xi}} \ = \ 0 ~.
\end{align}
In the classical field spaces of $h_{\chi,\xi}$, there are three minima: (1) The A point being the original point, which preserves the EW and $SU(2)_L \times SU(2)_R$ symmetries; (2) The B point being the electroweak vacuum; (3) and the C point where one have the symmetry breaking of the $SU(2)_L \times SU(2)_R\to SU(2)_V$.
\begin{align}
&A~point~:~h_{\phi} \to 0~,~h_{\xi} \to 0\; ;\nonumber\\
&B~point~:~h_{\phi} \to\nu_{\phi}~,~h_{\xi} \to \nu_{\xi}\;;\nonumber\\
&C~point~:~h_{\phi} \to 0~,~h_{\xi} \to \frac{-3 \mu_2+\sqrt{-12m_2^2 \lambda_2-4m_2^2 \lambda_3+9\mu_2^2}}{4(3\lambda_2+\lambda_3)}\;.\label{eqvac}
\end{align}
Here, we note that the $\mu_2$ parameter is crucial for the existence of the C point.
As shown in Fig.~\ref{fig:PTpattern}, as the temperature of the Universe cools down, the symmetry breaking process may occurs along the direction of $A\to B$ by one-step or $A\to C \to B$ by two-steps. The two correspond to one-step or two step phase transitions depending on the vacuum structure and potential height as will be explored in the following paragraph.

\label{sec:VSA}

The scalar potential of the EW vacuum should be the lowest one and the scalar potential of the original point is the maximal one of the three. The three scalar potentials $V_0(A),~V_0(B),~V_0(C)$ are,
\begin{align}
V_0(A) =&~0\;, \nonumber\\
V_0(B) =& - \lambda_1 \nu_\phi^4 - 3 \nu_\xi^3 (\mu_2 + (3\lambda_2+\lambda_3)\nu_\xi) - \frac{3}{8} \nu_\xi(\mu_1+4(2 \lambda4+\lambda5)\nu_\xi)\nu_\phi^2\;,\nonumber\\
V_0(C) =& -\frac{3}{256 \nu_{\xi}  (3 \lambda_2+\lambda_3)^3}( \textit{F} -3 \mu_2)^2 (\mu_2 (24 \nu_{\xi} ^2 (3 \lambda_2+\lambda_3)-2 \nu_{\xi}\textit{F})\nonumber\\
&+(3 \lambda_2+\lambda_3) (16 \nu_{\xi} ^3 (3 \lambda_2+\lambda_3)+\nu_{\phi}^2 (4 \nu_{\xi}  (2 \lambda_4+\lambda_5)+\mu_1))\;,\nonumber\\
&+6 \mu_2^2 \nu_{\xi} )\;, \\
\rm{where}\nonumber\\
\textit{\it{F}}=&\bigg(\frac{\nu_{\phi} ^2 (3 \lambda_2+\lambda_3) (4 \nu_{\xi}  (2 \lambda_4+\lambda_5)+\mu_1)}{\nu_{\xi} }+(4 \nu_{\xi}  (3 \lambda_2+\lambda_3)+3 \mu_2)^2\bigg)^{1/2}\;.
\end{align}
The two step phase transition might happen only when $-12m_2^2 \lambda_2-4m_2^2 \lambda_3+9\mu_2^2 \geq 0$ and $V_0(A)>V_0(C)>V_0(B)$ (with $\Delta V_0(AC)>0,\Delta V_0(CB)>0$). Meanwhile, the one-step phase transition would take place when $V_0(A)>V_0(B)$ ($\Delta V_0(AB)>0$) and $-12m_2^2 \lambda_2-4m_2^2 \lambda_3+9\mu_2^2 < 0$.  The potential differences are given as,
\begin{eqnarray}
\Delta V_0({AC})&\equiv&V_0(A)-V_0(C) \nonumber\\
&=& \frac{3}{256 \nu_\xi  (3 \lambda_2+\lambda_3)^3}( \textit{F} -3 \mu_2)^2 (\mu_2 (24 \nu_\xi ^2 (3 \lambda_2+\lambda_3)\nonumber\\
&-&2 \nu_\xi\textit{F})+(3 \lambda_2+\lambda_3) (16 \nu \xi ^3 (3 \lambda_2+\lambda_3)\nonumber\\
&+&\nu_\phi ^2 (4 \nu_\xi  (2 \lambda_4+\lambda_5)+\mu_1))\nonumber\\
&+&6 \mu_2^2 \nu_\xi) \\
\Delta V_0({CB})&\equiv&V_0(C)-V_0(B) \nonumber\\
&=& -\frac{3}{256 \nu_\xi  (3 \lambda_2+\lambda_3)^3}( \textit{F} -3 \mu_2)^2 (\mu_2 (24 \nu_\xi ^2 (3 \lambda_2+\lambda_3)\nonumber\\
&-&2 \nu_\xi\textit{F})+(3 \lambda_2+\lambda_3) (16 \nu_\xi ^3 (3 \lambda_2+\lambda_3)\nonumber\\
&+&\nu_\phi ^2 (4 \nu_\xi  (2 \lambda_4+\lambda_5)+\mu_1))\nonumber\\
&+&6 \mu_2^2 \nu_\xi)+\lambda_1 \nu_\phi^4 + 3 \nu_\xi^3 (\mu_2 + (3\lambda_2+\lambda_3)\nu_\xi)\nonumber\\
&+&\frac{3}{8} \nu_\xi(\mu_1+4(2 \lambda_4+\lambda_5)\nu_\xi)\nu_\phi^2\\
\Delta V_0({AB})&\equiv& V_0(A)-V_0(B) \nonumber\\
&=&\lambda_1 \nu_\phi^4 + 3 \nu_\xi^3 (\mu_2 + (3\lambda_2+\lambda_3)\nu_\xi) + \frac{3}{8} \nu_\xi(\mu_1+4(2 \lambda_4 \nonumber\\
&+&\lambda_5)\nu_\xi)\nu_\phi^2\;.
\end{eqnarray}\label{eq:potential}

\begin{figure*}[!htp]
\begin{center}
\includegraphics[width=0.4\textwidth]{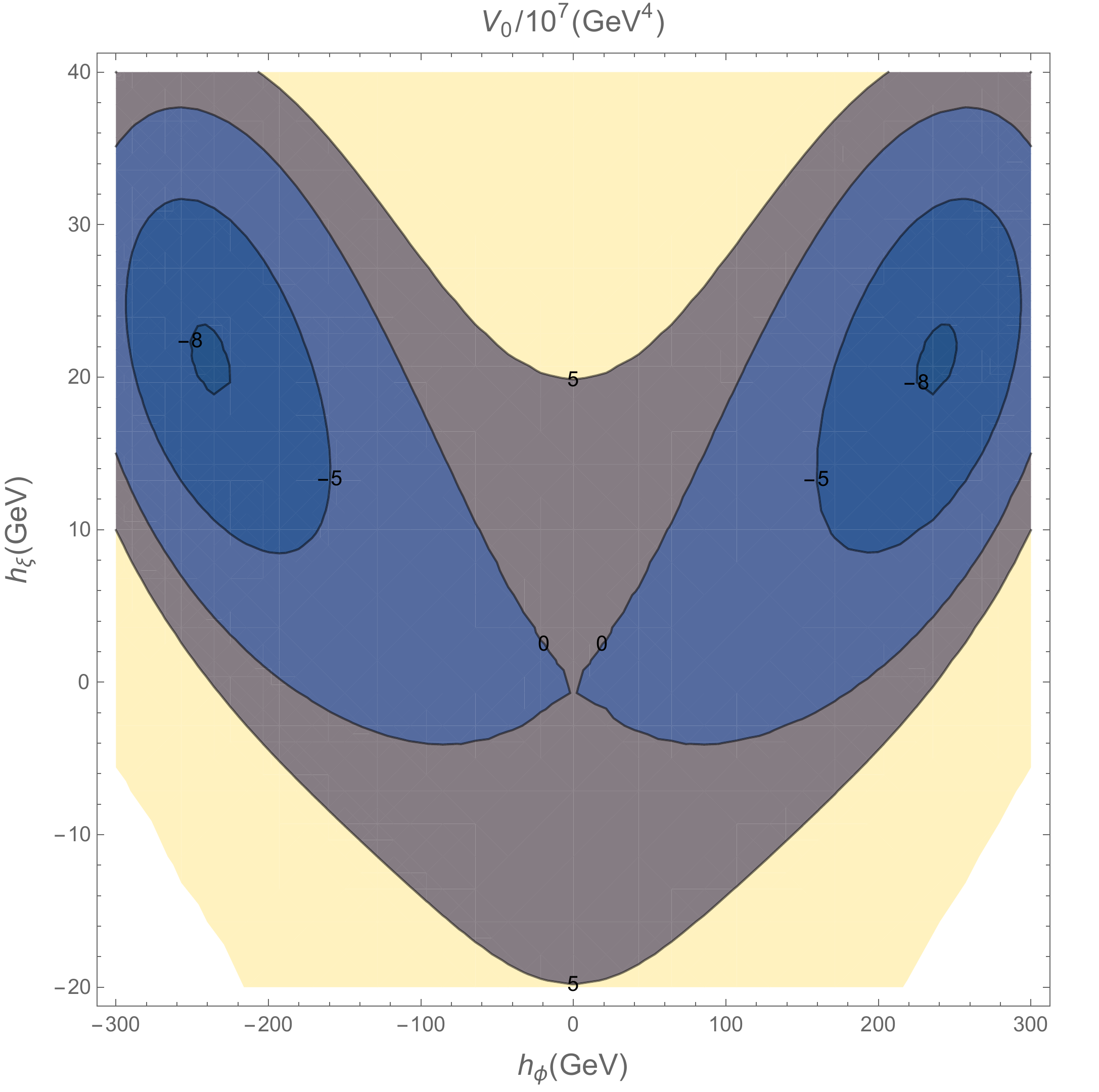}
\includegraphics[width=0.4\textwidth]{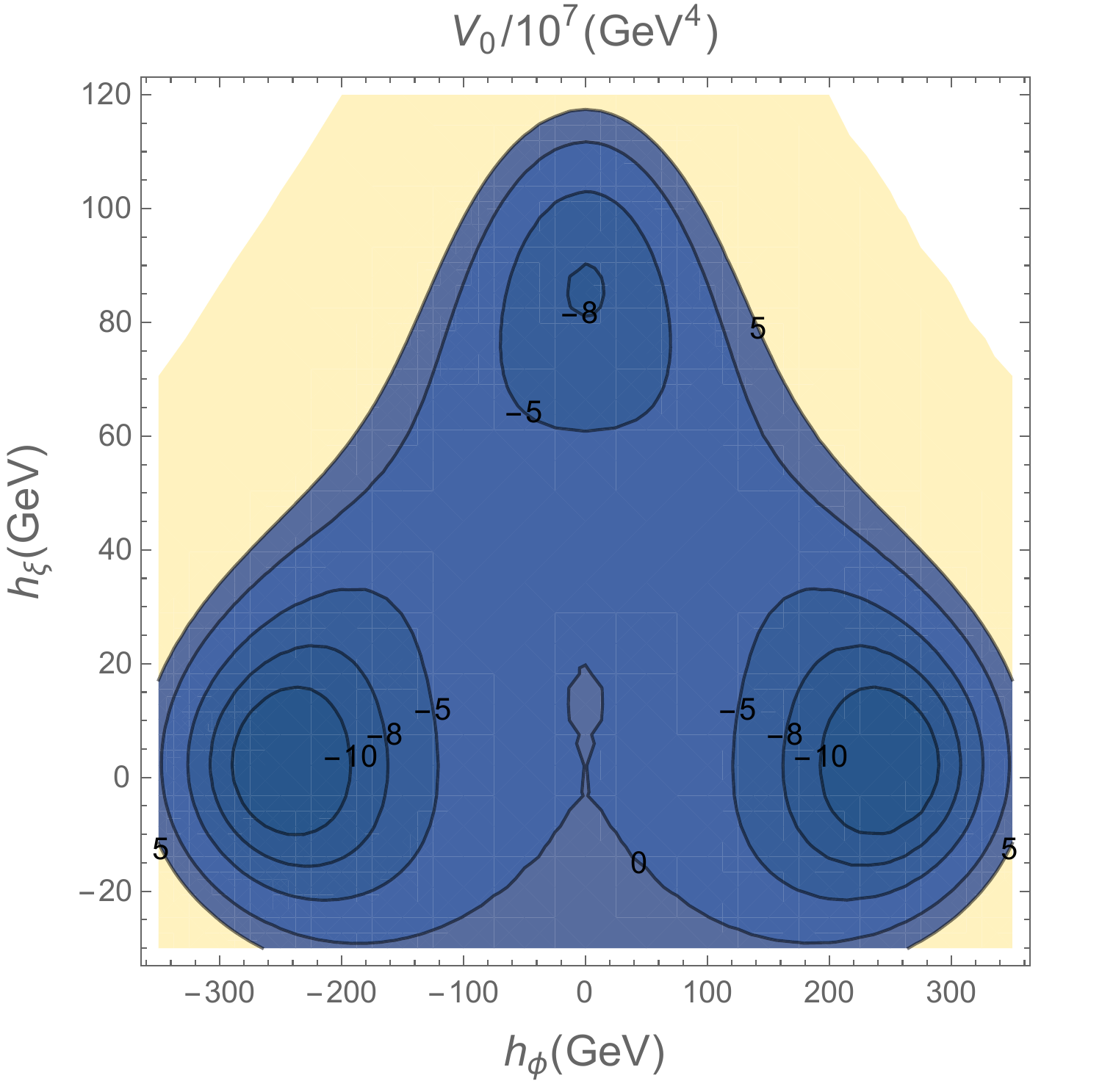}
\end{center}
\caption{The contours of $V_{0}$ (in units of $\rm{GeV}^4$) in $h_{\phi}$ - $h_{\xi}$ (in units of $\rm{GeV}$) plane. With the left plot parameters being: $\lambda_1 = 0.043$, $\lambda_2 = 0.933$, $\lambda_3 = -0.863$, $\lambda_4 = 0.321 $, $\lambda_5 = 1.267$, $\theta_H = 0.241$, $v = 246$GeV, $\mu_1=-289.785$GeV,$\mu_2=-2.928$GeV.  With the right plot parameters being $\lambda_1 = 0.0322$, $\lambda_2 = 1.069$, $\lambda_3 = -1.025$, $\lambda_4 = 0.706 $, $\lambda_5 = -0.311$, $\theta_H = 0.026$, $\mu_1=-11.128$GeV, $\mu_2=-143.970$GeV.}\label{treepotential}
\end{figure*}

Depending on the zero temperature vacuum structure and potential height, one may have one-step or two-step phase transition. Two-step phase transition may occur when a local minimum exists as shown in the right panel of the Fig.~\ref{treepotential}, and for the left panel plot case there could be a one-step phase transition.

\subsection{Phase Transition patterns}
\label{sec:php}

Utilizing the gauge invariant approach~\cite{Patel:2011th,Chao:2017vrq,Bian:2018mkl,Bian:2018bxr}, the finite temperature potential adopted for the study of phase transition behavior in the GM model is given by,
\begin{eqnarray}
V_{T}&=&V_{0}+\frac{ c_{\phi} T^2}{2}h_{\phi}^2 +\frac{ c_{\xi}T^2}{2}h_{\xi}^2+\frac{c_{\chi} T^2}{2}h_{\chi}^2\;,\label{eq:fP}
\end{eqnarray}
where the zero temperature potential $V_{0}$ is given by the Eq.~(\ref{eq:Vtree}) and the finite temperature corrections are calculated as,
 \begin{align}\label{cphichi}
c_{\phi}&=\frac{3 g^2 }{16}+\frac{g'^2 }{16}+\lambda_1+\frac{3 \lambda _4 }{4}+\frac{1}{4}  y_t^2 \sec ^2\theta_H\;,\nonumber\\
c_{\xi}&=\frac{g^2}{2}+\frac{11 \lambda _2 }{6}+\frac{7 \lambda_3 }{6}+\frac{\lambda_4 }{3}\;,\nonumber\\
c_{\chi}&=\frac{g^2 }{2}+\frac{g'^2 }{4}+\frac{11\lambda _2 }{6}+\frac{7 \lambda_3 }{6}+\frac{\lambda _4 }{3}\;.
\end{align}
Here, the $g,g',y_t$ are gauge couplings and top Yukawa coupling respectively.
We note that, with the high temperature approximation, the tadpole term involving $T^2 h_{\xi,\chi}$ that would make the symmetry break at high temperature does not appear as in the singlet model~\cite{Profumo:2014opa,Espinosa:2011ax} due to the global symmetry of $SU(2)_L\times SU(2)_R$ being imposed to the tree level potential. 
After substituting $h_\chi\to \sqrt{2} h_\xi$ in $V_T$ of Eq.~(\ref{eq:fP}), we obtain the finite temperature potential used for the phase transition study.
Due to the rich vacuum structures of the potential at finite temperature, there can be one-step or two-step phase transitions depending on whether the symmetry breaking of $SU(2)_L\times SU(2)_R\to SU(2)_V$ occurs earlier than the EW symmetry breaking. 
Ref.~\cite{Profumo:2014opa} indicates that realization of the one step phase transition requires
\begin{align}
c_{\phi,\chi,\xi}>0,~8 c_\phi+(c_\chi+2c_\xi)\tan^2\theta_H(T)&>0  \;,~
8m_1^2+3m_2^2\tan^2\theta_H(T)  <0\;,
\end{align}
where $\tan\theta_H(T)= 2\sqrt{2}h_\xi/h_\phi$ at finite temperature, the correspondence between the $\sin\theta_H(T_C)$ and $\sin\theta_H$ for both one-step and two-step SFOEWPT are given in Fig.~\ref{fig:sHT}. The $m_{1,2}$ here are given by the minimization of the tree level potential at the EW minimum, see Eq.~(\ref{m1m2}). Through which, the above conditions implicitly relay on the cubic Higgs coupling $\mu_1,\mu_2$.
For two step pattern phase transition, the T-dependence location of the minima can be described by the curves of~\cite{Espinosa:2011ax},  
\begin{align}
\frac{dD^2_{h_\phi}(h_\xi)}{dT^2}=-\frac{c_\phi}{\lambda_1}\;,
\frac{dD^2_{h_\xi}(h_\xi)}{dT^2}=-\frac{8h_\xi(c_\xi+2c_\chi)}{3(4(2\lambda_4+\lambda_5)h_\xi+\mu_1)}\;.
\end{align}
Following Ref.~\cite{Espinosa:2011ax}, 
to make the EW broken minimum (B point) the deepest one of the three vacua (A,B, and C point) for $T<T_C$, one needs the condition of $d(V_T(C)-V_T(B))/dT^2>0$. In turn, one has
\begin{align}
c_\xi (h_\xi^2(B)-h_\xi^2(C))+2c_\phi (h_\xi^2(B)-h_\xi^2(C))+c_\phi h_\phi^2(B)>0\;.
\end{align} 
where the B and C in the parentheses stand for the classical value of the fields ($h_{\phi,\xi}(T)$) at the EW breaking vacuum (around B point of Fig.~\ref{fig:PTpattern}) and $SU(2)_L\times SU(2)_R\to SU(2)_V$ vacuum ( around C points of Fig.~\ref{fig:PTpattern}). 
To ensure the B point being the global minimum at zero temperature, one needs the additional vacuum structure conditions being analyzed in Sec.~\ref{sec:VSA}. 
The $\mu_1$ parameter is useful for setting the tree level potential barrier for the phase transition, with the the term of $\mu_1h_\xi h_\phi^2$ in Eq.~(\ref{eq:Vtree}).
The $\mu_2$ parameter, as studied previously in Sec.~\ref{sec:VSA}, is another key parameter to ensure the possibility to have two-step phase transition with the existence of the C point of Eq.~(\ref{eqvac}). Our study will demonstrate that a typical region of $\mu_2$ is necessary for a two-step SFOEWPT to occur.

Usually, the one-step SFOEWPT requires relatively large cubic Higgs coupling, which may suffer from perturbativity problem of the model at high scale~\cite{Cheng:2018ajh,Cheng:2018axr}. The mixing between the SM Higgs and the additional Higgs can be easily tested by current and the future collider searches~\cite{Cheng:2018axr,Chen:2017qcz,Arkani-Hamed:2015vfh,Curtin:2014jma}, and therefore test the possibility to obtain the one-step SFOEWPT. Due to the typical vacuum structure, the two-step SFOEWPT can occur with much lower cubic Higgs coupling, see Ref.~\cite{Jiang:2015cwa,Grzadkowski:2018nbc} for the complex singlet model scenario which have a similar vacuum structure with the custodial symmetry conserving GM model. For completeness, we study both one-step and two-step SFOEWPT in this work. The SFOEWPT condition being $v_C/T_C\equiv \sqrt{h_\phi(T)^2+8 h_\xi(T)^2}/T_C\geq 1$ is adopted in this work with the $h_{\phi,\xi}(T)$ and $T_C$ evaluated as follows.

\subsubsection{One-step phase transition}
When the temperature of the Universe drops to the critical temperature with the Universe expands, two degenerate vacua (A and B points) occur with a potential barrier structure, which can be expressed as:
\begin{align}
&V_T(0,0,{T_C}) = V_T(h_{\phi}^B,h_{\xi}^B,{T_C})\; ,\nonumber\\
&\frac{{dV_T({h_\phi },{h_\xi },{T_C})}}{{d{h_\phi }}}{|_{{h_\phi } = {h_\phi^B },{h_\xi} = {h_\xi^B }}} = 0 \;, \frac{{dV_T({h_\phi },{h_\xi },{T_C})}}{{d{h_\xi }}}{|_{{h_\phi } = {h_\phi^B },{h_\xi} = {h_\xi^B }}} = 0\;,
\end{align}
through which, critical temperature and critical field value can be obtained. Here, we note that, to ensure two degenerate vacua occur the following constrains also should be satisfied:
$M_1P_1 - {N_1^2} > 0\;,M_1 > 0$, where
\begin{align}
\frac{{{d^2}V_T({h_\phi },{h_\xi },{T_C})}}{{dh_\phi ^2}}{|_{{h_\phi } = {h_\phi^B },{h_\xi } = {h_\xi^B }}} \equiv M_1 \;,
\frac{{{d^2}V_T({h_\phi },{h_\xi },{T_C})}}{{d{h_\phi }d{h_\xi }}}{|_{{h_\phi} = {h_\phi^B},{h_\xi} = {h_\xi^B}}} \equiv N_1\;, \nonumber\\
\frac{{{d^2}V_T({h_\phi },{h_\xi },{T_C})}}{{dh_\xi ^2}}{|_{{h_\phi } = {h_\phi^B },{h_\xi } = {h_\xi^B }}} \equiv P_1 \;.
\end{align}
The $h_{\phi,\xi}^B$ locates around $v_{\phi,\xi}$ at finite temperature $T_C$.

\subsubsection{Two-step phase transition}
When the temperature of the Universe drops to the critical temperature with the Universe expands, two degenerate vacua (B and C points) occur with a potential barrier structure, which can be expressed as:
\begin{align}
&V_T(0,h_{\xi}^C,{T_C}) = V(h_{\phi}^B,h_{\xi}^B,{T_C}) \;,\nonumber\\
&\frac{{dV_T({h_\phi },{h_\xi },{T_C})}}{{d{h_\phi }}}{|_{{h_\phi } = h_{\phi}^B,{h_\xi } = h_{\xi}^B}} = 0\;,
\frac{{dV_T({h_\phi },{h_\xi },{T_C})}}{{d{h_\xi }}}{|_{{h_\phi } = h_{\phi}^B,{h_\xi } = h_{\xi}^B}} = 0 \;,
\frac{{dV_T(0,{h_\xi },{T_C})}}{{d{h_\xi }}}{|_{{h_\xi } = h_{\xi}^C}} = 0 \;.
\end{align}
Through which, the critical temperature and critical field value can be obtained. As the one-step case, to ensure two degenerate vacua occur the 
following condition needs to be fulfilled:  $ M_2P_2 - {N_2^2} > 0,M_2 > 0 \;,
\frac{{{d^2}V_T(0,{h_\xi },{T_C})}}{{dh_\xi ^2}}{|_{{h_\xi} = h_{\xi}^C}} > 0\;$, with
\begin{align}
&\frac{{{d^2}V_T({h_\phi },{h_\xi },{T_C})}}{{dh_\phi ^2}}{|_{{h_\phi } = h_{\phi}^B,{h_\xi} = h_{\xi}^B}} \equiv M_2\;, \frac{{{d^2}V_T({h_\phi },{h_\xi },{T_C})}}{{d{h_\phi }d{h_\xi }}}{|_{{h_\phi} = h_{\phi}^B,{h_\xi} = h_{\xi}^B}} \equiv N_{2}\;, \nonumber\\
&\frac{{{d^2}V_T({h_\phi },{h_\xi },{T_C})}}{{dh_\xi ^2}}{|_{{h_\phi} = h_{\phi}^B,{h_\xi}= h_{\xi}^B}} \equiv P_{2}\;.
\end{align}
Here, the $h_{\phi,\xi}^B$ locates around $v_{\phi,\xi}$, and the $h_\xi^C$ locates around $h_\xi$ given in Eq.~(\ref{eqvac}) at finite temperature $T_C$.

\section{Phase transition patterns and the Higgs phenomenological prospects}
\label{sec:ptph}

We first study the general feature of the SFOEWPT in the GW model, which will reveal the relation between the triplet VEV and the SFOEWPT condition after considering the current and the projected experimental constraints: a lower magnitude of the triplet VEV is favored by both one-step and two-step SFOEWPT. The low mass $m_{h_5}$ benchmark is subsequently studied where small $\sin\theta_H$ is accompanied with $m_{h_5}<200$ GeV to be covered by future collider searches.

\subsection{The phase transition features in the GM model}

To consider theoretical and experimental constraints, we employ the GMCalc~\cite{Hartling:2014xma} to generate parameters of the GM model. Within these parameter spaces,
we perform phase transition analysis with the approach explored in Sec.~\ref{sec:ewptm}.

\begin{figure}[!htp]
\begin{center}
\includegraphics[width=0.3\textwidth]{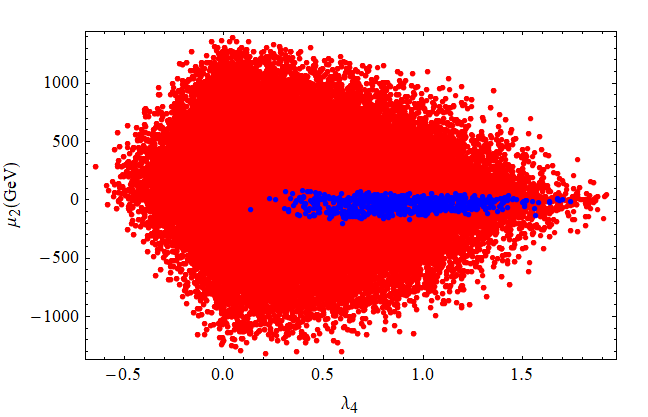}
\includegraphics[width=0.3\textwidth]{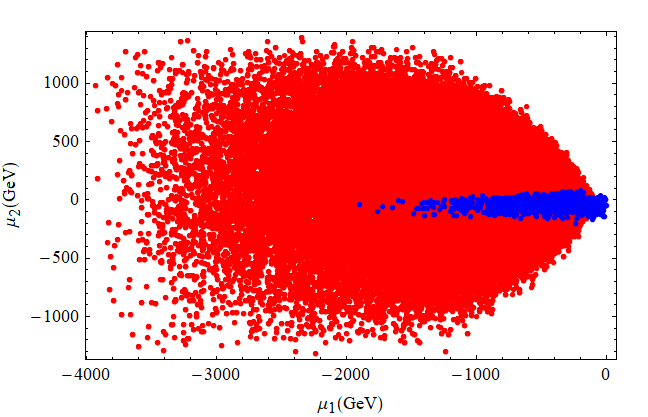}
\includegraphics[width=0.3\textwidth]{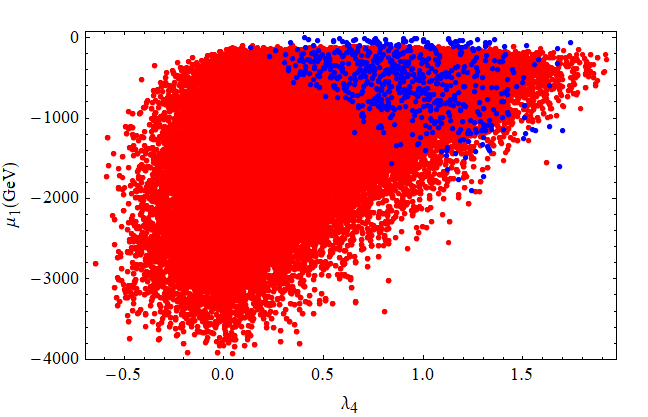}
\end{center}
\caption{ The SFOEWPT viable points in the parameter spaces of $\mu_2$ and $\lambda_4$(left), $\mu_2$ and $\mu_1$(middle), and $\mu_1$ and $\lambda_4$(right). The blue and red points satisfy the two step phase transition and one-step SFOEWPT conditions. }\label{fig:mu1}
\end{figure}

We first investigate how does the SFOEWPT relay on the cubic scalar couplings and the quartic scalar couplings.
Fig.~\ref{fig:mu1} depicts that one-step SFOEWPT usually requires a larger cubic scalar coupling of $\mu_1$, and the two-step SFOEWPT can occur with a relatively lower magnitude of $\mu_1$. This can trace back to the analysis of the phase transition patterns conditions in the Sec.~\ref{sec:php}. One can find that the one-step SFOEWPT can occur with both negative and positive $\lambda_4$, and the two-step SFOEWPT
can only occur in the parameter spaces with a positive $\lambda_4$.
The two-step SFOEWPT can occur with much lower magnitude of the cubic coupling $\mu_2$ in comparison with the one-step SFOEWPT, that reconfirms the vacuum structure analysis in Sec.~\ref{sec:ewptm}.

\begin{figure}[!htp]
\begin{center}
\includegraphics[width=0.4\textwidth]{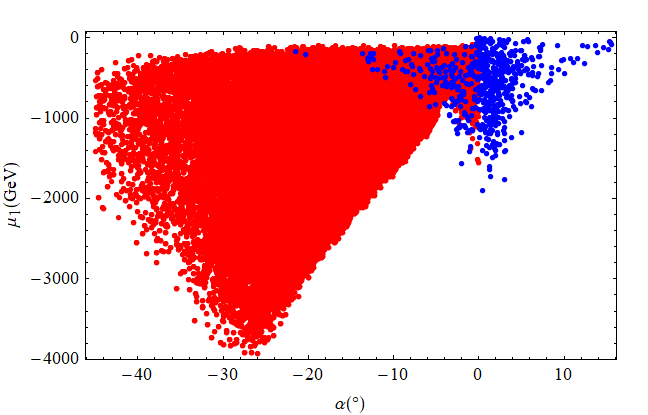}
\includegraphics[width=0.4\textwidth]{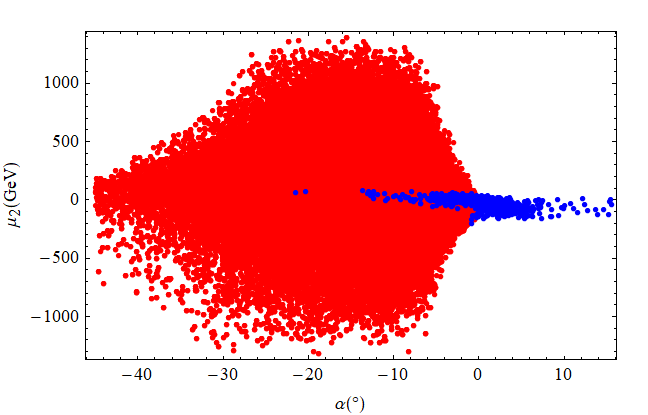}
\includegraphics[width=0.4\textwidth]{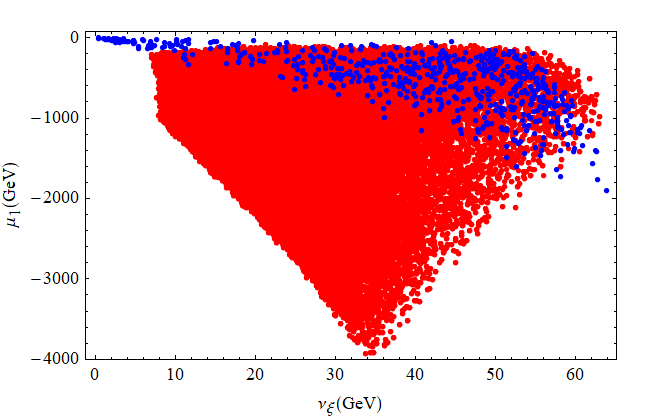}
\includegraphics[width=0.4\textwidth]{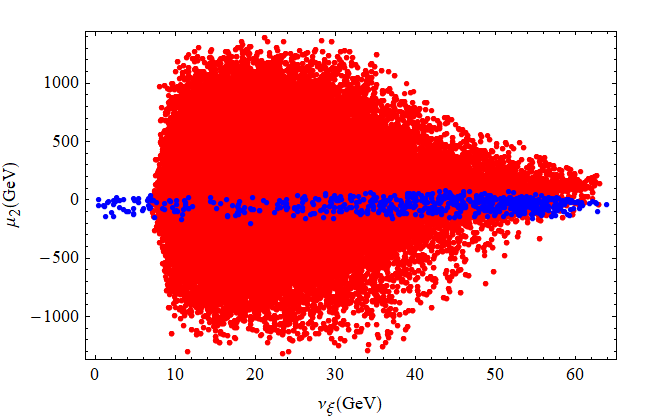}
\end{center}
\caption{ The SFOEWPT viable points in the parameter spaces of $\alpha$ and $\mu_1$(upper-left), $\alpha$ and $\mu_2$(upper-right), $\nu_\xi$ and $\mu_1$(below-left) and $\nu_\xi$ and $\mu_2$(below-right). The blue and red points satisfy the two step phase transition and one-step SFOEWPT conditions. }\label{fig:mu2}
\end{figure}

To reveal the relation between the phase transition and the mixing among the $h_{\phi,\chi,\xi}$ (the mixing angle $\alpha$), we plot the SFOEWPT allowed points in $\alpha$-$\mu_{1,2}$ plane.
In order to make clear how does the phase transition relay on the location of the $B$ point for one-step and two-step phase transition, we also show the SFOEWPT valid points in $\nu_\xi$-$\mu_{1,2}$ plane.
The top-left plot of the Fig.~\ref{fig:mu2} indicates that for a larger magnitude of $|\mu_1|$ a larger mixing angle of $\alpha$ is necessary for the one-step SFOEWPT, mostly $\alpha<0$. Meanwhile, the two step SFOEWPT can occur with a much higher probability around $\alpha\sim0$ (both $\alpha<0$ and $\alpha>0$ are allowed) with small mixing between the light and heavy Higgs or $h_\phi$ and $h_{\chi,\xi}$, and a smaller $|\alpha|$ is accompanied with a larger $\mu_1$. The top-right panel of Fig.~\ref{fig:mu2} reflects the same information as the middle panel of Fig.~\ref{fig:mu1} because the $\alpha$ is characterized by $\mu_1$.
The bottom-left panel indicates that the two-step SFOEWPT can occur with a larger $|\mu_1|$ in the parameter regions with a larger $v_\xi$, which means a larger $\tan\theta_H$ . In addition to the negative value of $\alpha$, positive $\alpha$ angel can also lead to two-step SFOEWPT. The bottom-right plot demonstrates that the occurrence possibility of the one-step SFOEWPT decreases as $v_\xi$ (or $\sin\theta_H$) increases, while the two-step SFOEWPT almost occur with the same probability with relatively lower magnitude of $\mu_2$.

\begin{figure}[!htp]
\begin{center}
\includegraphics[width=0.385\textwidth]{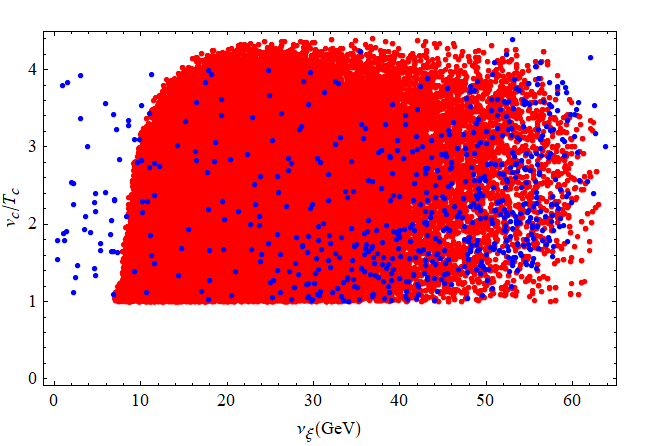}
\includegraphics[width=0.4\textwidth]{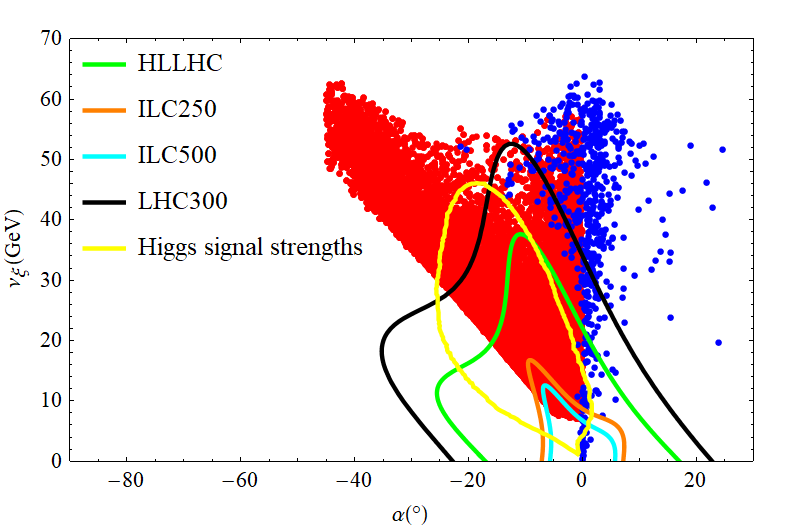}
\end{center}
\caption{ Left:The $v_C/T_C$ as a function of $v_{\xi}$ for one step(red) and two-step phase transition(blue);   Right: The $v_C/T_C>1$ viable points in the plane of $\alpha$-$v_{\xi}$ for one step and two-step SFOEWPT.}\label{fig:vctcxi}
\end{figure}

To make a better understanding how does the phase transition relay on the VEV of the triplet, and furthermore the phase transition patterns, we perform the survey of the relation between the SFOEWPT condition and $v_\xi$ in Fig.~\ref{fig:vctcxi}.  The left plot shows that SFOEWPT can occur in the parameter regions of $v_\xi< 60$ GeV.  One can read the strength of the phase transition in the plot, which is found to be almost $v_C/T_C\leq 4.5$ for one and two step cases. It seems that the possibility to reach one-step SFOEWPT is much higher than the two-step one, and there is a tendency that $v_C/T_C$ increases with $v_\xi$ for the two-step case.
The Ref.~\cite{Chiang:2015amq,Li:2017daq,Chiang:2018cgb} performed the study of the constraints on the $v_\xi$-$\alpha$ parameter spaces of the GM model. Their results mostly bound the parameter spaces to a negative $\alpha<0$ due to the LHC Higgs signal strength constraints.
The projected 14 TeV LHC with 300 fb$^{-1}$, HL-LHC with 3000 fb$^{-1}$, and ILC at 250 and 500 GeV can also probe the parameter spaces. The right plot of fig.~\ref{fig:vctcxi} demonstrates that the one-step SFOEWPT can occur for an increasing $\nu_\xi$ as $\alpha$ increases. The 13 TeV LHC Higgs signal strength fitted by Ref.~\cite{Chiang:2018cgb} has been adopted here to restrict the GM model parameter, which set severe bounds on the parameter spaces of the two-step SFOEWPT: $\alpha\approx 0$ with $v_\xi\leq 20$ GeV, which corresponds to $\sin\theta_H<0.23$.

\begin{figure*}[!htp]
\begin{center}
\includegraphics[width=0.4\textwidth]{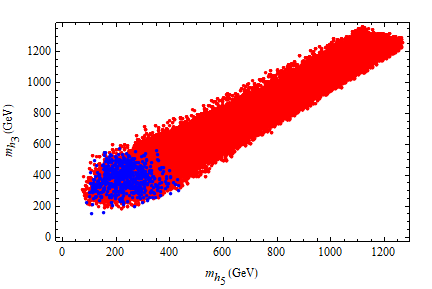}
\includegraphics[width=0.385\textwidth]{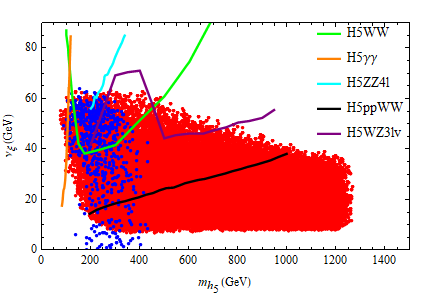}
\end{center}
\caption{Left: The $v_C/T_C>1$ viable points in the plane of $m_{h_3}-m_{h_5}$for one-step (red) and two-step (blue) phase transitions; Right: The $v_C/T_C>1$ viable points in the plane of $\sin\theta_H-m_{h_5}$ for one-step (red) and two-step (blue) phase transition with the upper limits of $\sin\theta_H$ from different experiments.   }\label{shmh5}
\end{figure*}

To understand the relation between the SFOEWPT condition and the charged Higgs mass $m_{h_5}$, we plot Fig.~\ref{shmh5}.
The left panel of Fig.\ref{shmh5} shows the SFOEWPT valid point in the parameter spaces of $m_{h_3}-m_{h_5}$.  The two-step SFOEWPT favors 200 GeV$<m_{h_3}<$600 GeV together with 100 GeV$<m_{h_5}<400$ GeV. The right panel of Fig.\ref{shmh5} indicates one can have only two-step SFOEWPT for $\sin\theta_H<0.1$, one-step SFOEWPT requires slightly larger $m_{h_5}$ for smaller $\sin\theta_H$ or $v_\xi$, for the two-step case one has relatively smaller $v_\xi$ and $\sin\theta_H$ accompany with lower value of $m_{h_5}$.
We impose collider search bounds in the plane of $\sin\theta_H-m_{h_5}$.
Curves with different colors are the constraints from current experiment searches for heavy scalars, where the green, cyan and black lines are the upper bounds from $WZ$ channel~\cite{Aaboud:2018ohp,Sirunyan:2017sbn,CMS:2018ysc}, the yellow and orange lines are the bounds from $WW$ channel ~\cite{Aaboud:2017gsl,Aaboud:2017fgj}, the pink and brown lines are the bounds from $ZZ$ channel~\cite{Aaboud:2017rel,Aaboud:2017itg}. The purple line which is the limit from same-sign $WW$ searches of the doubly charged scalar is the strongest constraints in this parameter space.
The same-sign $WW$ channel ruled out the SFOEWPT validated parameter spaces (including both one-step and two-step) with relatively large $\sin\theta_H$ increasing with $m_{h_5}$. This motivates the study of the two-step SFOEWPT in the parameter spaces with low magnitudes of $\sin\theta_H$ and $m_{h_5}$.

We summarize the relation between the SFOEWPT and Higgs phenomenology explored in this section as follows:
\begin{itemize}

\item {\it On the trilinear scalar couplings}
We first review the information we got on the relation between the SFOEWPT condition and the trilinear scalar couplings, $\mu_{1,2}$, that constitute a main contribution to the triple Higgs couplings, as shown in Eq.~(\ref{eq:3h}) and will be discussed later.
In comparison with the two-step SFOEWPT, relatively higher magnitude of $\mu_1$ is required for one-step SFOEWPT. Lower value of $\mu_2$ (close to 0) is required to realize the two-step SFOEWPT.

\item {\it The SFOEWPT valid parameter spaces confronting with the experimental constraints}
The right panel of the Fig.~\ref{fig:vctcxi} indicates that the 13 TeV LHC signal strength measurements would restrict the possibility to obtain two-step SFOEWPT in the relatively narrow region $\alpha\sim 0$. The projected electron-positron colliders would constrain the magnitude of $v_\xi$ to even lower value, and therefore narrow down the possibility to obtain the SFOEWPT through one-step and two-step patterns. The right panel of Fig.~\ref{shmh5} shows that the larger $\sin\theta_H$ would be excluded by the same-sign WW boson channel search at 13 TeV LHC, which means that the possibility to reach SFOEWPT would be bounded to the parameter spaces with small $\sin\theta_H$ and small $\alpha$.
\end{itemize}

\subsection{ On triple Higgs couplings and the SFOEWPT condition}

\begin{figure}[!htp]
\begin{center}
\includegraphics[width=0.3\textwidth]{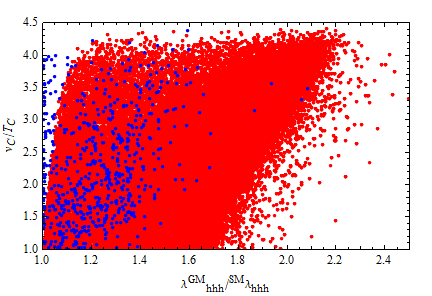}
\includegraphics[width=0.3\textwidth]{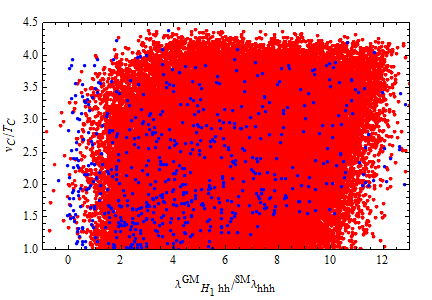}
\includegraphics[width=0.3\textwidth]{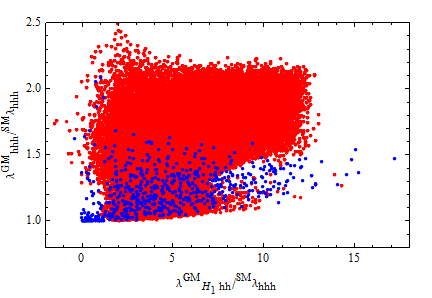}
\includegraphics[width=0.3\textwidth]{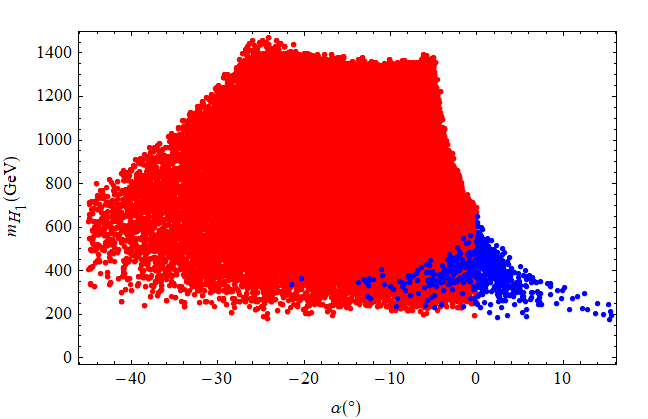}
\includegraphics[width=0.3\textwidth]{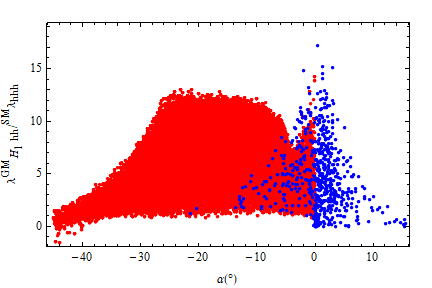}
\includegraphics[width=0.3\textwidth]{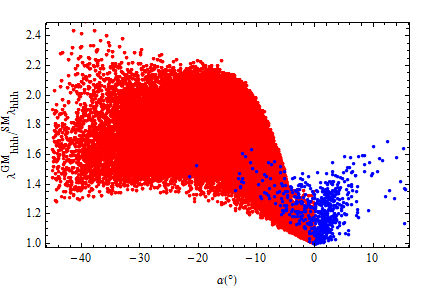}
\end{center}
\caption{The phase transition strength as a function of the triplet Higgs coupling (top-left: $\lambda_{hhh}$, top-middle: $\lambda_{H_1hh}$) for one step(red points) and two step(blue points) SFOEWPT, the relation between $\lambda_{hhh}$ and $\lambda_{H_1hh}$ is also shown in the top-right panel. In the bottom panels, the relations between $m_{H_1}$(left), $\lambda_{H_1hh}$(middle), $\lambda_{hhh}$(right) and $\alpha$ are shown for one- and two-step SFOEWPT.}\label{onehhh}
\end{figure}

To reveal the relation between the SFOEWPT and the triple Higgs coupling to be searched at $e^+e^-$ and $pp$ colliders, we plot Fig.~\ref{onehhh}.
The top-left and top-middle plots of the Fig.~\ref{onehhh} illustrate that the magnitude of the triple Higgs couplings($\lambda_{hhh}^{GM}$ and $\lambda_{H_1 hh}^{GM}$) grows with the increase of the strength of phase transitions for both one-step and two-step SFOEWPTs, which means larger deviation from the SM case can lead to a larger $v_C/T_C$. The deviation of triple Higgs couplings can be probed through the Higgs associated production process at lepton colliders and Higgs pair search at hadron colliders (LHC, FCC-hh, and SPPC), we refer to Ref.~\cite{Li:2017daq} and Ref.~\cite{Chang:2017niy} for recent studies.
The top-right panel depicts that, the two-step phase transition can occur in the parameter spaces with a smaller $\lambda_{hhh}^{GM}$ and a larger $\lambda_{H_1 hh}^{GM}$.
In this situation, the constraint from Higgs pair productions is usually more powerful, see Ref.~\cite{Chang:2017niy} for a recent study. The bottom-left plot shows that for the mixing angel of $\alpha\sim 0$, the two-step phase transition can occur with a largest value of $m_{H_1}\sim $600 GeV. For the parameter spaces with a positive value of $\alpha$, the phase transition can be two-step SFOEWPT with a relatively lower mass of the CP-even Higgs (200 GeV $<m_{H_1}<$600 GeV). Furthermore, the bottom-middle (bottom-right) plot shows that
the triple Higgs coupling $\lambda^{GM}_{H_1 hh}$ ($\lambda^{GM}_{hhh}$) increases (decreases) with the decrease of $|\alpha|$.
One can probe the two-step SFOEWPT valid parameter spaces with the Higgs pair searches at LHC, Fcc-hh and SPPC, we left the detailed studies to the future study.

The interaction strength between the Higgs and SM fermions and gauge bosons are:
\begin{eqnarray}
&&g_{hf\bar{f}}=\cos\alpha/\cos\theta_H g^{SM}_{hf\bar{f}}\,,~g_{hVV}=(\cos\alpha\cos\theta_H-\sqrt{\frac{8}{3}}\sin\alpha\sin\theta_H) g^{SM}_{hf\bar{f}} \;,\nonumber\\
&&
g_{H_1 f\bar{f}}=\sin\alpha/\cos\theta_H g^{SM}_{hf\bar{f}}\, ,~ g_{H_1 VV}=(\sin\alpha\cos\theta_H+\sqrt{\frac{8}{3}}\cos\alpha\sin\theta_H )g^{SM}_{hVV}\;.\nonumber
\end{eqnarray}
In the scenario with small $\alpha$ and small $\theta_H$, the $g_{hf\bar{f},hVV}$ close to the SM case, $g_{H_1f\bar{f},H_1 VV}$ are suppressed.
For small mixing angle $\alpha$ bounded by LHC, HL-LHC, and ILC, the triple Higgs coupling given in Eq.~(\ref{eq:3h}) recast as
\begin{eqnarray}
&&\lambda^{GM}_{h hh}\approx \lambda_{hhh}^{SM}-\frac{3\sqrt{3}}{2}\mu_1\sin\alpha\;,\nonumber\\
&&\lambda^{GM}_{H_1 hh}\approx \sqrt{6}(\lambda_4 v\sin\theta_H+\frac{\lambda_5}{2} v \sin\theta_H-\frac{\mu_1}{2})+(\lambda_{hhh}^{SM}-8\lambda_4 v \cos\theta_H-4\lambda_5v\cos\theta_H)\sin\alpha\;.\nonumber
\end{eqnarray}
The first formula implicit that a large $\mu_1$, corresponding to most parameter spaces of the one-step SFOEWPT, can lead to a large deviation of the $\lambda^{GM}_{hhh}$ from the $\lambda^{SM}_{hhh}$. Thus the probe of the triple Higgs coupling of $\lambda^{GM}_{hhh}$ through the $e^+e^-$ colliders should be able to test the parameter regions where the one-step SFOEWPT can be realized.
For small $\theta_H$, one has $g_{H_1 f\bar{f}} \times \lambda^{GM}_{H_1 hh}\approx -\frac{\sqrt{6}}{2}\sin\alpha/\cos\theta_H  \mu_1\times g_{hf\bar{f}}^{SM} $ and $g_{H_1 VV}\times  \lambda^{GM}_{H_1 hh}\approx \frac{\sqrt{6}}{2}\mu_1\times g_{H_1 VV}$.
That means that the Higgs pair searches of the $H_1$ utilizing to test the two-step SFOEWPT parameter regions (at LHC, FCC-hh, and SPPC) for $\alpha,\theta_H\to 0$ is characterized by the parameter $\mu_1$.

\subsection{The ``H5plane"}
\begin{figure*}[!htp]
\begin{center}
\includegraphics[width=0.4\textwidth]{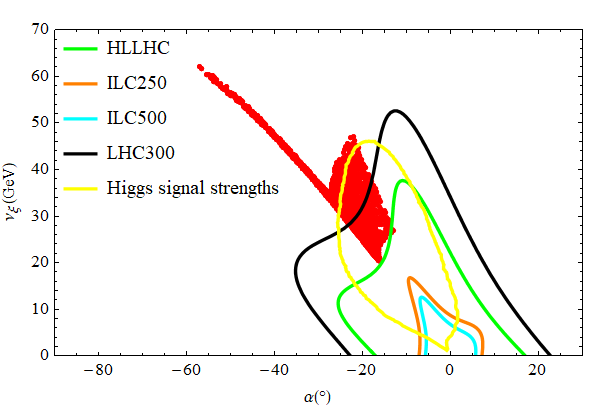}
\includegraphics[width=0.4\textwidth]{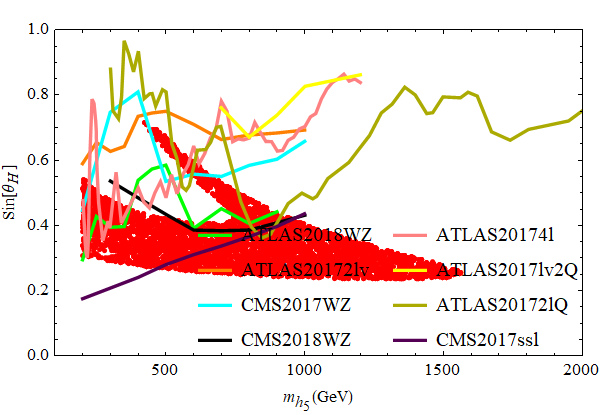}
\end{center}
\caption{Left:The $v_C/T_C>1$ viable points in the plane of $\alpha-v_\xi$ for one-step phase transition; Right: The $v_C/T_C>1$ viable points in the plane of $\sin\theta_H-m_{h_5}$ for one-step phase transition.   }\label{h5plane}
\end{figure*}

At last, we comment on the phase transition behaviors within the ``H5plane" benchmarks developed by the LHC Higgs Cross Section Working Group~\cite{deFlorian:2016spz}.
The Ref.~\cite{Logan:2017jpr} studied the ``H5plane" benchmark scenarios, where the regions with smaller $m_{h_5}\leq 200$ GeV and $\alpha\geq 0$ (see Fig.~\ref{fig:vctcxi} and Fig.\ref{shmh5}) that are crucial for the two-step SFOEWPT are not covered. That results in losing of the possibility of the two-step SFOEWPT, see the Fig.~\ref{h5plane}. In the following section, we would go beyond this benchmark and study the phase transition in the scenario including low mass $m_{h_5}\leq 200$ GeV case.

\subsection{Low mass charged Higgs benchmarks}

As illustrated in the previous section, the current same-sign $WW$ channel searches of doubly charged Higgs at the LHC preclude the possibility to obtain SFOEWPT in the parameter spaces with large $\sin\theta_H$ along with the increase of $m_{h_5}$. Therefore, we consider the electroweak phase transition in one particular benchmark scenario, the low mass fermiophobic charged scalar considered in~Ref.~\cite{Logan:2018wtm}. In this case, the quintuple is the lightest scalar multiplet, the usual searching channels involving gauge bosons are suppressed by $\sin\theta_H$ which is chosen to be much less than 1, as well as by the phase space where the quintuple is light and below the diboson threshold. This leaves only four physical input parameters that are relevant in this benchmark, which can be chosen as follows: two parameters $m_{h_5}$ and $\delta m^2$ that control the mass spectrum of the heavy Higgs bosons, and two parameters $\sin\theta_H$ and $\mu_2$ that control the couplings.
In particular, we adapt the definition of the benchmark from Ref.~\cite{Logan:2018wtm} with small modifications as listed in Tab.~\ref{tab:benchmark}.
The parameters are chosen to accommodate low $\sin\theta_H$ as indicated in Ref.~\cite{Logan:2018wtm}.

\begin{table}[!bt]
\centering
\begin{tabular}{l|ll}
\hline\hline
Variable Parameters & \multicolumn{2}{c}{Other Parameters}\\
\hline
$m_{h_5}\in [100,500]$ GeV & $m_{h_3}^2 = m_{h_5}^2 + \delta m^2$ & $\delta m^2 = (100+\frac{m_{h_5}}{2})^2$ GeV$^2$  \\

$\mu_2 \in [-300,300]$ GeV & $m_{H_1}^2 = m_{h_5}^2 + \frac{3}{2}\delta m^2 + \kappa_H \sin\theta_H^2 \nu^2$& $\kappa_H = 1-2\left(\frac{m_{h_5}}{250\text{ GeV}}\right)^2$\\

$\sin\theta_H \in [0.0,0.5]$& $\mu_1 = -\left[\frac{\sqrt{2}}{\nu}\left(m_{h_5}^2+\frac{3}{2}\delta m^2\right) - 3\mu_2\sin\theta_H + \kappa_{\lambda_3}\nu \sin\theta_H^2\right]\sin\theta_H$ &$\kappa_{\lambda_3}=\frac{1}{2}-\frac{m_{h_5}}{150\text{ GeV}}$ \\
$\sin\alpha\in [-\kappa_\alpha,\kappa_\alpha]$ & & $\kappa_\alpha =  \frac{m_{h_5}}{1000\text{ GeV}}\sin\theta_H$\\
\hline
\end{tabular}
\caption{The parameter choice in low mass benchmark.}
\label{tab:benchmark}
\end{table}


\begin{figure*}[!htp]
\begin{center}
\includegraphics[width=0.4\textwidth]{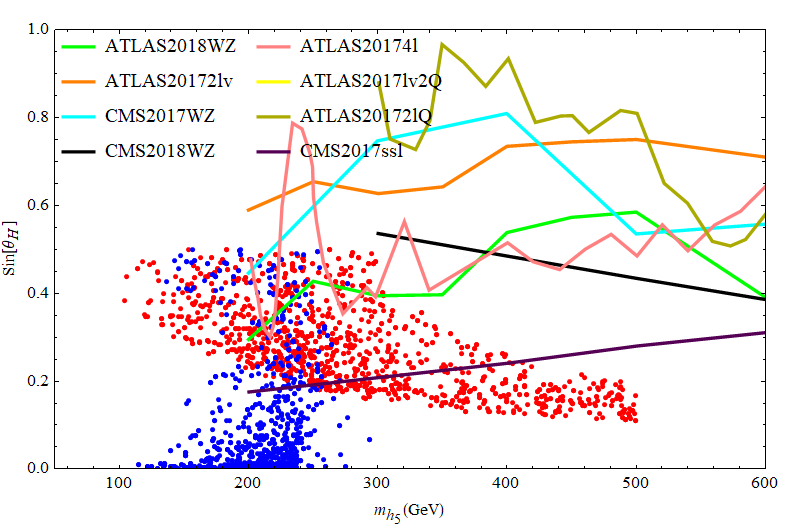}
\end{center}
\caption{The $v_c/T_c>1$ viable points in the $m_{h_5}$-$\sin\theta_H$ plane for one step(red) and two-step(blue) phase transition for the low mass $m_{h_5}$ scenario.  }\label{lowmassshmh5}
\end{figure*}

In Fig.~\ref{lowmassshmh5}, we show SFOEWPT viable points in the parameter space of $\sin\theta_H$ and $m_{h_5}$. The blue and red points satisfy the two-step and one-step SFOEWPT conditions respectively. Several current experimental constraints are also imposed in this plane represented by the lines with different colors. By construction, the low mass benchmark can evade most searching channels involving gauge bosons ($WZ$ channel (green, black and cyan lines)~\cite{Aaboud:2018ohp,Sirunyan:2017sbn,CMS:2018ysc}, $WW$ channel (yellow and orange lines)~\cite{Aaboud:2017gsl,Aaboud:2017fgj} and $ZZ$ channel (pink and brown lines)~\cite{Aaboud:2017rel,Aaboud:2017itg}). The same-sign $WW$ search from doubly charged scalar (purple line)~\cite{Sirunyan:2017ret} has the sensitivity down to about 0.17 for $\sin\theta_H$ at $m_{h_5}\sim 200$ GeV. In this case, the same-sign $WW$ search is still the strong constraint and can explore large viable parameter spaces of one-step SFOEWPT for $m_{h_5} > 200$ GeV. The figure also indicates that extending the search of same-sign $WW$ down to lower mass region could explore most one-step SFOEWPT parameter space. While two-step viable points can extend to lower $\sin\theta_H$ region, which is difficult for $WZ/WW/ZZ$ channels that are suppressed by $\sin\theta_H$.

\begin{figure*}[!htp]
\begin{center}
\includegraphics[width=0.4\textwidth]{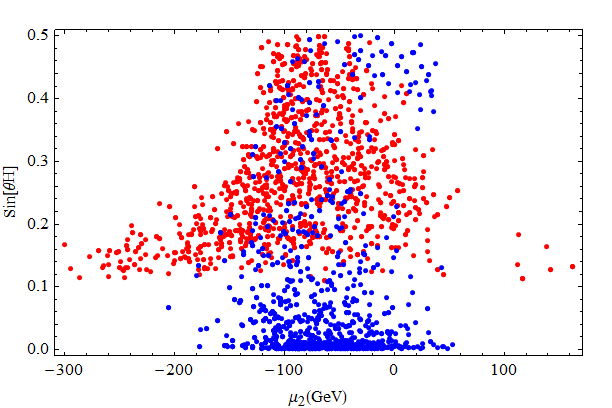}
\end{center}
\caption{The $v_c/T_c>1$ viable points in the plane of $\sin\theta_H$ over $\mu_2$ for one step(red) and two-step(blue) phase transition for the low mass $m_{h_5}$ scenario. }\label{lowmassshmh5sh}
\end{figure*}

\begin{figure*}[!htp]
\begin{center}
\includegraphics[width=0.4\textwidth]{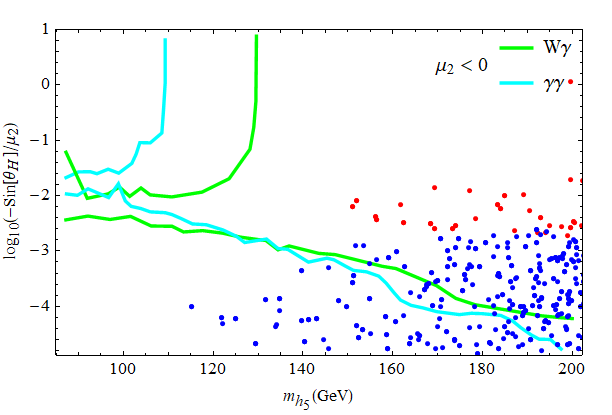}
\end{center}
\caption{The constraints of $W\gamma$ (green line) and $\gamma\gamma$ (cyan line) channels in low mass benchmark. The region to the left of the contours can be excluded by corresponding searches. 
}
\label{fig:wgamma-gammagamma}
\end{figure*}

For this low $\sin\theta_H$ and low mass region, as indicated in Ref.~\cite{Logan:2018wtm}, the most sensitive channels are those loop-induced channels ($\gamma\gamma$ and $W\gamma$ etc). The dominant contributions to these channels come from the triple scalar couplings which is proportional to $\mu_2$ and is not suppressed by $\sin\theta_H$. In the Fig.~\ref{lowmassshmh5sh} we show the SFOEWPT valid points including one-step and two-step in the $\mu_2$-$\sin\theta_H$ plane, the SFOEWPT mostly occurs for a negative $\mu_2$. We can see that lower value for $\sin\theta_H$ is accompanied by larger range of $\mu_2$ for one-step SFOEWPT valid point, which means the triple scalar coupling is generally large in this region and leads to large partial width of these loop-induced channels.

In the low mass region, when $WZ/WW/ZZ$ channels are suppressed by the phase space as well as by $\sin\theta_H$, the loop-induced channels will dominate. We impose the constraints from $W\gamma$ channel from Ref.~\cite{Logan:2018wtm} as well as the diphoton search from 8 TeV ATLAS~\cite{Aad:2014ioa} in this low mass region, the results are shown in Fig.~\ref{fig:wgamma-gammagamma} where the red and blue points present the one- and two-step SFOEWPT viable points as before. 
The green line and cyan line are the constraints from $W\gamma$ and $\gamma\gamma$ searches respectively.
We find that the loop induced decay channel has the sensitivity for the two-step viable parameter space in low mass region. Further improvement (such as higher luminosity, extending to lower mass region for $\gamma\gamma$ channel) in these searches will explore most of the two-step SFOEWPT parameter space.

\section{Conclusions}
\label{sec:conc}

In this work, we study the EWPT in the frame work of GM model. The EWPT can be SFOEWPT through one-step or two-step as the Universe cools down. In comparison with the one-step SFOEWPT, the two-step SFOEWPT can occur with a relatively smaller trilinear couplings between the Higgs doublet and the triplet scalars. The contribution of the triplets to the electroweak symmetry breaking is crucial for the SFOEWPT and the Higgs phenomenology, which is characterized by the parameter $\sin\theta_H$ or $v_\xi$.
The bounds from LHC, HL-LHC and the projected ILC constraints in the $\alpha$-$\nu_\xi$ plane can lower the possibility to realize SFOEWPT, especially the two-step SFOEWPT. The 13 TeV LHC Higgs signal strength measurements limit the two-step SFOEWPT valid parameter spaces to a rather small region with $\alpha\sim 0$. The current same-sign WW search performed at the 13 TeV LHC ruled out a lot parameter spaces to realize SFOEWPT including both one-step and two-step.

The Higgs pairs search would be able to search the two-step SFOEWPT when one have a nonnegative mixing angle $\alpha\geq 0$ accompanied with a lower mass of $m_{H_1}$. For much smaller triplet VEV (with the $\sin\theta_H<0.1$) and lower quintuple mass, the phase transition could be two-step SFOEWPT rather than one-step. The $WW$, $W\gamma$ and $\gamma\gamma$ searches for the $m_{h_5}<200$ GeV parameter regions can probe the two-step SFOEWPT within the GM model in this low mass and small $\sin\theta_H$ region.

At last, we note that the custodial symmetry in the Higgs potential prohibits the CP violation in the GM model. To address the BAU with the EWBG mechanism in the GM model, one may need to introduce a tinny custodial symmetry breaking without violate the $\rho$ parameter constraints to include CP violation phases or introduce additional CP violations through high dimensional operators.

\section*{Acknowledgement}

We thank Zhi-Long Han and Bin Li for helpful discussions on the collider bounds on the GM model. We are grateful to Chengwei Chiang and Kei YagYu for helpful communications on the custodial symmetry of the GM model. The work of LGB is Supported by the National Natural Science Foundation of China (under grant No.11605016 and No.11647307), Basic Science Research Program through the National Research Foundation of Korea (NRF) funded by the Ministry of Education, Science and Technology (NRF-2016R1A2B4008759), and Korea Research Fellowship Program through the National Research Foundation of Korea (NRF) funded by the Ministry of Science and ICT (2017H1D3A1A01014046). The work of Y.C.W. is partially supported by
the Natural Sciences and Engineering Research Council of Canada.

\appendix
\section{GMcalc convention vs our convention}

For the reader's convenience, we give the comparison of the convention adopted by GMCalc and this work in Table.~\ref{conv}.

\begin{table}[!htp]
\caption{GMcalc convention vs our convention}\label{conv}
\begin{center}
\begin{tabular}{c c c }
\hline
~~~~~~~~~~~GMCalc~~~~~~~~~~& ~~~~~~~~~This work~~~~~~~~~ \\
\hline
    $\mu_2$  &$m_1$ \\
    $\mu_3$ & $m_2$  \\
  $\lambda_1$ & $\lambda_1$  \\
  $\lambda_4$ & $\lambda_2$  \\
  $\lambda_3$ & $\lambda_3$  \\
  $\lambda_2$ & $\lambda_4$  \\
  $-\lambda_5$ & $\lambda_5$  \\
    $-M_1$ & $\mu_1$  \\
    $-M_2$ & $\mu_2$  \\
\hline
\end{tabular}
\end{center}
\end{table}

\section{The triplet Higgs couplings}

The trilinear Higgs couplings can be read:
\begin{eqnarray}\label{eq:3h}
g_{hhh}&=&24 \cos\alpha^3 \lambda_1 \nu_\phi+6 \cos\alpha \sin\alpha^2 \nu_\phi (2 \lambda_4+\lambda_5)
+\frac{3}{2} \sqrt{3} \cos\alpha^2 \sin\alpha  (4 \nu_{\xi} (-2 \lambda_4-\lambda_5)-\mu_1)\nonumber\\
&-& 4 \sqrt{3} \sin\alpha^3 (\mu_2+2 \nu_{\xi} (3 \lambda_2+\lambda_3))\;,\\
g_{H_1hh}&=&24\lambda_1 \cos\alpha^2 \sin\alpha \nu_{\phi}+2[\sqrt{3}\cos\alpha \nu_{\xi}(3 \cos\alpha^2-2)
+\sin\alpha\nu_{\phi}(1-3\cos\alpha^2)]\nonumber\\
& \times &(2\lambda_4+\lambda_5)+8\sqrt{3}\cos\alpha \sin\alpha^2 \nu_{\xi}(\lambda_3+3\lambda_2)
+\frac{\sqrt{3}}{2}\mu_1 \cos\alpha(3\cos\alpha^2-2)\nonumber\\
&+&4\sqrt{3}\mu_2 \cos\alpha \sin\alpha^2\;.
\end{eqnarray}

\section{On custodial symmetry}

\begin{figure*}[!htp]
\begin{center}
\includegraphics[width=0.6\textwidth]{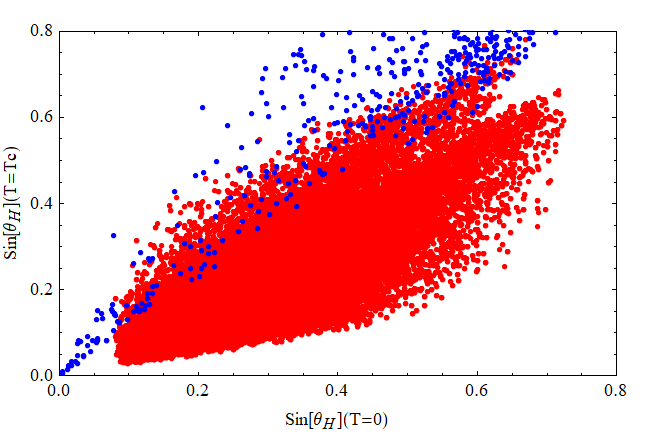}
\end{center}
\caption{The $\sin\theta_H$ as $T=T_C$ and $T=0$ for one-step and two-step SFOEWPT. 
}
\label{fig:sHT}
\end{figure*}

In the previous sections, we focus on the custodial symmetry preserving case. Previous studies of Ref.~\cite{Blasi:2017xmc,Keeshan:2018ypw} shows that
the custodial symmetry that are preserved at tree level can be explicitly break by loop effects of the $U(1)_Y$ hyper- charge gauge interaction, the custodial symmetry breaking could be probed at future $e^+e^-$ colliders.
At zero temperature, the $\rho$ parameter is described by,
\begin{eqnarray}
\rho=\frac{v_\phi^2+4 v_\chi^2+4 v_\chi^2}{v_\phi^2+8 v_\chi^2}=\frac{v^2}{v^2+4(v_\chi^2-v_\xi^2)}\;.
\end{eqnarray}
Thought the custodial symmetry study is beyond this work, we briefly list useful formula for the finite temperature study.
At finite temperature, with the $\rho$ parameter can be parametrized as,
\begin{eqnarray}
\rho(T)=\frac{h_\phi^2(T)+4 h_\chi(T)^2+4 h_\xi(T)^2}{h_\phi(T)^2+8 h_\chi(T)^2}\;,
\end{eqnarray}
In the first stage phase transition of the two-step phase transition, one have null $h_\phi(T)$ with the $\rho(T)\equiv 1$.
While, for the second stage phase transition of the two-step phase transition, one have,
\begin{eqnarray}
m_{h_\xi}^B&=& \frac{1}{4 h_{\xi} }(-16 \lambda_2 h_{\xi}^3-16 \lambda_3 h_{\xi}^3-8 \lambda_4 h_{\xi}   h_{\phi} ^2- \mu_1 h_{\phi}^2-16 \lambda_2 h_{\xi}   h_{\chi}^2-12 \mu_2 h_{\chi}^2-2 \sqrt{2} \lambda_5 h_{\chi} h_{\phi}^2)\;,\nonumber\\
m_{h_\chi}^B&=&\frac{1}{4 h_{\chi} }(-2 \sqrt{2} \lambda_5 h_{\xi}  h_\phi ^2-\sqrt{2} \mu_1 h_{\phi}^2-16 \lambda_2 h_{\chi}^3-8 \lambda_3 h_{\chi}^3-16 \lambda_2 h_{\xi}^2 h_{\chi} -8 \lambda_4 h_{\chi} h_{\phi} ^2-2\lambda_5 h_{\chi} h_{\phi}^2\nonumber\\
&-&24 \mu_2 h_{\xi} h_{\chi})\;\nonumber\\
m_{h_\phi}^B&=&\frac{1}{2} (-8 \lambda_1 h_{\phi}^2-4 \lambda_4 h_{\xi}^2-  \mu_1 h_\xi -4 \lambda_4 h_{\chi}^2 - \lambda_5 h_{\chi}^2-2 \sqrt{2} \lambda_5 h_{\xi}  h_{\chi} -\sqrt{2} \mu_1 h_{\chi})\;,
\end{eqnarray}
where
\begin{eqnarray}
m_{h_\xi} &=&m_2^2+ c'_\xi T^2 \;,\\
m_{h_\chi} &=&   m_2^2+c'_\chi T^2 \;,\\
m_{h_\phi} &=&  m_1^2 + c'_\phi T^2\;,
\end{eqnarray}
and
 \begin{eqnarray}
c'_\xi&=& \frac{g^2}{2} +  \frac{2\lambda_2}{3} +  \frac{\lambda_3}{2} +  \frac{\lambda_4}{12}\;,\\
c'_\chi&=& \frac{g^2}{2} +  \frac{g'^2 }{4} +  \frac{2\lambda_2}{3} +  \frac{\lambda_3}{4} +  \frac{\lambda_4}{12}+  \frac{\lambda_5}{48}\;,\\
c'_\phi&=&\frac{3g^2}{16} +  \frac{g'^2 }{16} +  \frac{\lambda_1}{16} +  \frac{\lambda_4}{6} +  \frac{\lambda_5}{48} +  \frac{1 }{4}y_t^2 \sec^2(\theta H)\;.
\end{eqnarray}
Suppose the condition of $h_\chi =\sqrt{2}h_\xi$ maintains as studied in this work, the following relations maintains,
\begin{eqnarray}
c_\xi T^2 &=& -\frac{1}{4 h_{\xi} } (8 c_\chi T^2 h_\xi +3 (16 {h_{\xi}}^3 (3 \lambda_2+\lambda_3)+\mu_1 h_{\phi}^2+24 \mu_2 h_\xi^2+4 h_{\xi}(h_{\phi}^2 (2 \lambda_4+\lambda_5)+m_2^2)))\;\nonumber\\
c_\phi T^2 &=& -4 \lambda_1 h_{\phi}^2 - m_1^2 - \frac{3}{2} h_{\xi}  (4 h_{\xi} \lambda_4 +2 \lambda_5 h_{\xi} + \mu_1)\;,
\end{eqnarray}
with $c_{\chi,\phi}$ being given in Eq. \ref{cphichi}. The correspondence of $\sin\theta_H(T)=\frac{2\sqrt{2}h_\xi(T)}{\sqrt{h_\phi(T)^2+8 h_\xi(T)^2}}$ at $T=T_C$ and $T=0$ is given in the Fig.~\ref{fig:sHT}, which indicates the match situation of the two.
After the temperature drops below $T_C$, the $\sin\theta_H(T_C)$ would evolve to be the $\sin\theta_H$ fininally (i.e., $\sin\theta_H(T=0)$) and one obtain the EW symmetry break vacuum with $SU(2)_V$ symmetry.  

\newpage

\bibliographystyle{unsrt}

\end{document}